\documentclass[reprint, amsmath, amssymb, aps]{revtex4-1}
\usepackage{amsmath}
\usepackage{amsfonts}
\usepackage{graphicx}
\usepackage{appendix}
\usepackage{dsfont}
\usepackage{amssymb}
\usepackage{bm}
\usepackage{xcolor}
\usepackage{ulem}
\usepackage{soul}
\usepackage{mathrsfs}
\usepackage{amsmath}
\usepackage{esint}%\usepackage{endfloat}
\usepackage{natbib}
\usepackage{nccmath}
\usepackage{array}
\newcommand{\beq}{\begin{equation}}
\newcommand{\eneq}{\end{equation}}
\newcommand{\be}{\begin{equation}}
\newcommand{\ee}{\end{equation}}
\newcommand{\bea}{\begin{eqnarray}}
\newcommand{\eea}{\end{eqnarray}}

\usepackage{multirow}
\usepackage{wasysym}

\begin{document}

\title{Lindblad master equation approach to the  dissipative quench dynamics of planar superconductors}
 
\author{Andrea Nava$^{(1)}$,   Carmine Antonio Perroni$^{(2)}$,   Reinhold Egger$^{(1)}$, Luca Lepori$^{(3)}$, 
and Domenico Giuliano$^{(4)}$}
\affiliation{
$^{(1)}$Institut f\"ur Theoretische Physik,
Heinrich-Heine-Universit\"at, 40225 D\"usseldorf, Germany \\
$^{(2)}$ Dipartimento di Fisica ``E. Pancini'' Complesso Universitario Monte S. Angelo Via Cintia, I-80126 Napoli, 
Italy and \\ CNR-SPIN, Complesso Universitario Monte S. Angelo Via Cintia, I-80126 Napoli, Italy and  \\
I.N.F.N., Sezione di Napoli, Complesso Universitario Monte S. Angelo Via Cintia, I-80126 Napoli, Italy \\
$^{(3)}$Dipartimento di Scienze Matematiche Fisiche e Informatiche Universit\`a  di Parma and\\
  INFN, Gruppo Collegato di Parma, Parco Area delle Scienze 7/A, 43124, Parma, Italy. \\
$^{(4)}$Dipartimento di Fisica, Universit\`a della Calabria Arcavacata di 
Rende I-87036, Cosenza, Italy and \\ I.N.F.N., Gruppo collegato di Cosenza 
Arcavacata di Rende I-87036, Cosenza, Italy \\}

\begin{abstract}

We employ the  Lindblad master equation method  to study the
 nonequilibrium dynamics following a parametric quench in the
 Hamiltonian   of an open, two-dimensional superconducting 
system  coupled to an  external bath.  Within our approach we show how,  in the open system, the 
dissipation works as  an effective stabilization mechanism in the time 
evolution of the system after the quench.   Eventually, we evidence how 
the mismatch between the phases corresponding to the initial and 
to the final  state of the system determines a dynamical phase 
transition  between the two distinct phases. Our method allows for fully characterizing 
the dynamical phase transition in an open system in several cases of physical relevance,  by means of a
combined study of the  time-dependent superconducting gap and of the fidelity between density matrices.

\end{abstract}
\date{\today}
\maketitle

\section{Introduction}
\label{intro}

Related to the  continuous developments of time-resolved spectroscopic investigation methods in many-particle systems, 
there has recently been an increasing interest in  
nonequilibrium  correlated systems. For instance, using time-dependent angle-resolved spectroscopy, it becomes possible to investigate 
the different time evolutions of quasiparticle states in a superconductor in different regions of the Brillouin zone, together 
with the corresponding effects on the dependence in time of the superconducting gap 
\cite{Graf2011,Smallwood2014,Peronaci2015}. Also, pertinently irradiating the system, it is possible to 
induce the onset of metastable transient states, with peculiar properties, sometimes completely different 
from the ones of the ``true'' asymptotic state   reached as the time $t \to \infty$ \cite{Caviglia2012,Nava2018}. 

There are at least two main issues that arise in studying the time evolution of nonequilibrium correlated systems. 
First of all, typically, such systems   are characterized by several different phases \cite{Lee2006}, 
often close to each other in energy. Knowing their transient dynamics allows for finding out 
  to which phase they flow, once prepared in a given state,  thus 
  recovering crucial information about their elementary  excitations  \cite{Andre2012,Sandri2015}. 
Also, controlling their time evolution allows for possibly stabilizing metastable phases, with novel, exotic physical properties, 
sometimes  rather different from the ones characterizing the equilibrium states \cite{Fu2014,Nava2018}. 
In addition, along their time evolution, it is possible, for the systems, to go through a dynamical phase 
transition (DPT), driven by the time $t$, between the initial state, in which they are
 prepared at  $t=0$, toward the final state, 
to which they evolve  as $t\to\infty$  \cite{Zvyagin2016,Heyl2018,Heyl2019}.

A widely implemented protocol to induce nonequilibrium dynamics in a many-electron system 
consists in preparing it  in the ground state of a specific Hamiltonian, in 
performing a sudden quench in some parameter(s) of the system Hamiltonian, and eventually 
in making the system evolve with the final (``after the quench'') Hamiltonian. 
In the specific case of a   superconducting electronic system, the protocol outlined above results 
in an effective  time dependence of the superconducting gap, which can be accounted for by means of  
a time-dependent generalization of the self-consistent mean-field
 (SCMF) approach  \cite{Peronaci2015,Mazza2017}.

In this paper, we define and study a procedure for inducing nonequilibrium dynamics in two-dimensional (2D)
 superconducting  systems, involving  two, or more than two, components
of the order parameter with different symmetry (such as, for instance, an $s$-wave and a $d$-wave 
component of the superconducting gap).  In analogy to Ref. \cite{Peronaci2015}, we set the nonequilibrium 
dynamics by quenching the interaction strength(s) of the corresponding model Hamiltonian. Eventually, we  
recover the time-dependent superconducting gap by systematically implementing self-consistency, at
any given time $t> 0$. In addition,  we employ the Lindblad master equation (LME) approach to the dissipative 
dynamics of the density matrix of the system 
\cite{Petruccione2002,Wilde2013,Nava2019,Manzano2020,Nava2022,Artiaco2023,Mazza2023}, to account for dissipation and 
damping effects  beyond the time-dependent SCMF approximation. Such effects are  related to the interaction among quasiparticles, as well
as to the   coupling between the quasiparticles and the fluctuations of the superconducting order parameter \cite{Cui2019}.
  In fact, we do not derive the LME,  rather we consider the most generic equations 
that can drive the system to thermal equilibrium. As detailed in  Ref. \cite{Petruccione2002}, this is a  standard approach,
 based on imposing the detailed balance condition and considering all the independent operators defined within the system’s Hilbert space that allow transitions between different system eigenstates.

Within the LME framework, we couple the system to an external bath, able to exchange energy and quasiparticles with the 
system itself. In doing so, we show how the relaxation dynamics induced by the coupling to 
 the bath naturally drives the superconductor toward its
asymptotic, stationary state. We 
conclude, therefore,  that the dissipation works as  an effective stabilization mechanism in the time 
evolution of the system after the quench. Eventually, the mismatch between the phases corresponding to the initial and 
to the asymptotic state of the superconductor can drive the system across a  real-time 
 DPT   between the two distinct phases  \cite{Zvyagin2016,Heyl2018,Heyl2019}. 
 
  In fact, while the SCMF approach is expected to be unable to capture the complex interplay of nearby phases in strongly correlated superconductors, such as, for instance,  cuprates  in their underdoped region, it still allows for effectively highlighting the physics of simple models, such as  the one we employ here \cite{Peronaci2015}. Moreover, we argue how, resorting to the LME approach, eventually allows for accounting for effects beyond the SCMF approximation, such as the interaction among quasiparticles, as well as the direct coupling between the quasiparticles and the fluctuations of the superconducting order parameter \cite{Cui2019}.

 DPTs typically arise in the time evolution of quantum systems  after a  parametric quench in  the system  Hamiltonian 
\cite{Heyl2013,Jurcevic2017,Schmied2019,Yuzbashyan2006b,Prufer2018,Yamamoto2021,Debashish2022,Debashish2023}.  In our specific case, 
in addition to looking at the   time dependence of the superconducting order parameter, we 
 approach the DPT by computing the fidelity ${\cal F}(t)$ between the initial state of the system, 
 $|\psi(0)\rangle$, and its state  at time $t$.  Indeed, differently from a closed system, where 
 a DPT is    typically investigated by 
looking at the singularities in the  Loschmidt echo ${\cal L} (t) = | \langle \psi (0) |\psi (t) \rangle|^2$ \cite{Heyl2019,Pollmann2010,Heyl2013,Abeling2016,Bhattacharya2017,Lang2018}, in open systems
  the  Loschmidt echo (as well as quantities related to it) is no longer applicable to monitor 
the DPT and it has to be substituted by some more appropriate quantities, such as the fidelity ${\cal F} (t)$ 
\cite{Heyl2019,Wu2022}.
 
 Although, in this paper, we focus on  a limited number of phase transitions,  the effectiveness of our method is 
 grounded on  its wide applicability to many different choices for the superconducting gap, 
 such as, for instance, the ones appropriate for 2D oxide superconductors 
 \cite{Biscaras2012, Scheurer2015,Perroni2019,Lepori2021}. Moreover, by looking at how the time dynamics
 of the system is affected by the choice of the actual values of the system parameters, we can
 in principle suggest how to tune the parameters of realistic devices so to realize phases with 
the desired properties, including a nontrivial topology \cite{shorter_paper}. Finally,  our approach allows, 
via a synoptic monitoring of the time-dependent superconducting gap,  the fidelity, and (in case of a topological DPT, which 
we address in Ref. \cite{shorter_paper})   the spin-Hall conductance, for a  comprehensive 
characterization of a  DPT.

Our paper is organized as follows:

\begin{itemize}

\item In Sec. \ref{model}, we  present  our general two-dimensional  lattice model Hamiltonian   for a planar superconductor, 
we  employ the SCMF  approximation to trade it for an effectively quadratic one, we 
 map out the different superconducting phases as a function of the interaction strengths, and  we introduce   
 the LME approach to the system coupled to the   bath.

\item In Sec. \ref{timevol},  we  discuss in detail the relaxation dynamics of our 
superconducting system  for different choices of the superconducting order parameter before, and after, the 
sudden change in the interaction strengths. 
 
\item In Sec. \ref{toprt}, we compute the fidelity and employ it to characterize a DPT.
   
\item In Sec. \ref{conclusions} , we discuss and summarize our results and present some possible further extensions of our work.

\item In the Appendixes, we present the technical details of our calculations.

\end{itemize}

\section{Model Hamiltonian and methods}
\label{model}

We now   present  our  lattice model Hamiltonian $H$ for  a planar superconductor. $H$
 encompasses   various  interaction terms    (on-site, nearest-neighbor, next-to-nearest neighbor), allowing for
various possible kinds of spin-singlet superconducting order parameters. We then employ the SCMF approximation
to recover the phase diagram of $H$ as a function of the different interaction strengths.   
Finally, we  present  the LME approach, 
which  describes the dynamics of the nonequilibrium system  coupled to 
the  bath.

 \subsection{Model Hamiltonian for the lattice planar superconductor}
 \label{mhlattice}
 
\begin{figure}
 \center
\includegraphics*[width=.9 \linewidth]{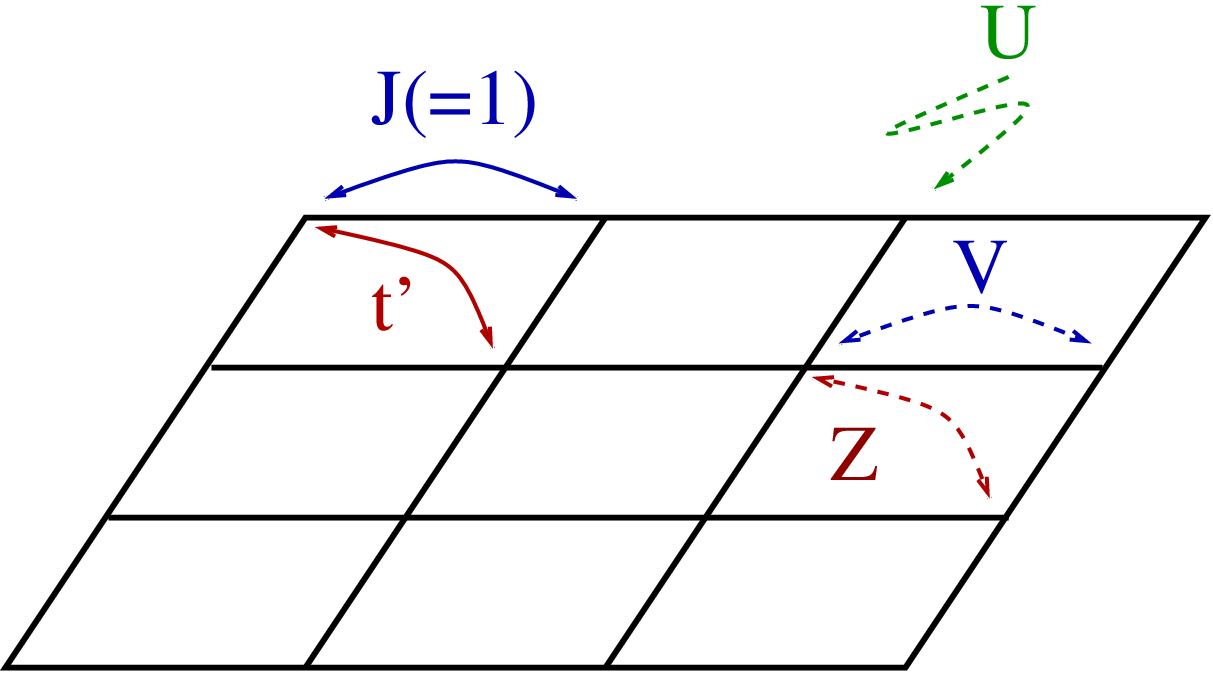}
\caption{Sketch of the square lattice  with   the various single-fermion hopping and interaction terms
in Eq.(\ref{eh.1a}): the NN  (solid  blue)  and the NNN (solid red) hopping terms,  
  the on-site (dashed green), the NN (dashed blue), and the NNN (dashed red) interaction terms. }
\label{fig_0}
\end{figure}

 Our main model Hamiltonian describes a system of interacting  spinful electrons, defined over a  2D  square lattice.
 The single-particle dispersion relation is determined by a nearest-neighbor (NN) hopping strength $J$ (which we 
 will use as our unit of energy, i.e., $J=1$), and 
 a next-to-nearest neighbor (NNN) hopping strength $t'$. In addition, we allow for   finite
 on-site,   NN   and   NNN  density-density  interactions,
 all in the spin-singlet channel,  
 with interaction strength respectively given by $U$, $V$ and $Z$.  
 Accordingly,   $H$ is   given by (see Fig.\ref{fig_0})

\begin{eqnarray}
&& H=-\mu \sum_{{\bf r} } \sum_\sigma c_{{\bf r},\sigma}^\dagger c_{{\bf r},\sigma} 
\label{eh.1a} \\
&& - \sum_{{\bf r},\hat{\delta}}\sum_\sigma
c_{{\bf r} +\hat{\delta},\sigma}^\dagger c_{{\bf r} ,\sigma} -t'\sum_{{\bf r},\hat{\delta}'}\sum_\sigma
c_{{\bf r}+\hat{\delta}',\sigma}^\dagger c_{{\bf r},\sigma} \nonumber \\
&&-U\sum_{{\bf r}} n_{{\bf r},\uparrow}n_{{\bf r},\downarrow} - \frac{V}{2}\sum_{{\bf r},\hat{\delta}} n_{{\bf r}} n_{{\bf r}+\hat{\delta}}
- \frac{Z}{2}\sum_{{\bf r},\hat{\delta}'} n_{{\bf r}} n_{{\bf r}+\hat{\delta}'} 
\;, \nonumber 
\end{eqnarray}
\noindent
with $c_{{\bf r},\sigma}, c_{{\bf r},\sigma}^\dagger $ being the annihilation and the creation operators for an electron
with spin $\sigma$ at site ${\bf r}$ of a square lattice and $\mu$ being the chemical potential. 
$c_{{\bf r},\sigma},c_{{\bf r},\sigma}^\dagger$ satisfy the canonical 
anticommutation relations $\{c_{{\bf r},\sigma},c_{{\bf r}',\sigma'}^\dagger \} = \delta_{{\bf r},{\bf r}'} \delta_{\sigma,\sigma'}$. 
The spin-polarized density operators in Eq.(\ref{eh.1a}) are defined as $n_{{\bf r},\sigma}=c_{{\bf r},\sigma}^\dagger c_{{\bf r},\sigma}$, while
$n_{{\bf r}}=\sum_\sigma n_{{\bf r},\sigma}$. In Eq.(\ref{eh.1a}) we have set the lattice constant to 1. 
$\hat{\delta}$ denotes a generic (unit length) vector connecting 
${\bf r}$ with the corresponding NN sites of the lattice, while  $\hat{\delta}'$ denotes a generic vector (of length $\sqrt{2}$), connecting 
${\bf r}$ with the corresponding NNN sites of the lattice. 
In the context of solid-state systems, the Hamiltonian $H$ in Eq.(\ref{eh.1a}) is a 
generalization of model Hamiltonians widely applied to describe high-$T_c$ superconductors 
\cite{Laughlin1998,Ghosh1999,Salkola1998,Ghosh2002}. Within alternative platforms, such as cold-atom condensates, 
optical realizations of systems effectively described by Hamiltonians similar to $H$ are nowadays within the reach 
of present technology \cite{Goldman2016}.

\noindent
 
In Appendix \ref{mhe}, we implement the SCMF approximation  to trade $H$  in 
Eq.(\ref{eh.1a}) for the corresponding mean-field,  quadratic (in the fermionic operators) Hamiltonian $H_{\rm MF}$, given by 

\beq
H_{\rm MF}=\sum_{{\bf k}}\sum_\sigma \xi_{\bf k} c_{{\bf k},\sigma}^\dagger c_{{\bf k},\sigma} - \sum_{\bf k} \{\Delta_{\bf k}c_{{\bf k},\uparrow}^\dagger 
c_{{\bf -k},\downarrow}^\dagger + {\rm h.c.}\} 
\;\;\; , 
\label{eh.2a}
\eneq
\noindent
with h.c. standing for Hermitean conjugate and 
with the single-fermion operators in momentum space, $c_{{\bf k},\sigma}$, related to the $c_{{\bf r},\sigma}$'s by means of 

\beq
c_{{\bf k},\sigma} = \frac{1}{\sqrt{N}} \sum_{\bf r} e^{-i{\bf k}\cdot {\bf r}} c_{{\bf r},\sigma}
\;\; , 
\label{eh.3a}
\eneq
\noindent
 $N$ being the number of lattice sites. Also, in Eq.(\ref{eh.2a}) we have set 

\begin{eqnarray}
\xi_{{\bf k}}&=&-2 [ \cos(k_x)+\cos(k_y)]-4t'\cos(k_x)\cos(k_y)-\mu \nonumber \\
\Delta_{\bf k}&=&\Delta_S+2\Delta_{x^2-y^2} \{\cos(k_x) -\cos(k_y)\}\nonumber \\
&-&
 4i\Delta_{xy}\sin(k_x)\sin(k_y) 
\;\;\;,
\label{eh.4a}
\end{eqnarray}
\noindent
with $\Delta_S,\Delta_{x^2-y^2},\Delta_{xy}$ respectively being equal to the s-wave, to the 
d-wave and to the id-wave components of the superconducting order parameters. As we show in 
Appendix \ref{mhe}, they are 
determined by the self-consistent equations 

\begin{eqnarray}
\Delta_S &=& \frac{U}{2N} \sum_{\bf k} \frac{\Delta_{\bf k}}{\epsilon_{\bf k}} \nonumber \\
\Delta_{x^2-y^2} &=& \frac{V}{4N} \sum_{\bf k} \frac{ [\cos (k_x ) - \cos (k_y ) ] \Delta_{\bf k}}{\epsilon_{\bf k}} \nonumber \\
\Delta_{xy} &=& \frac{iZ}{2N} \sum_{\bf k}  \frac{ \sin (k_x ) \sin (k_y )  \Delta_{\bf k}}{\epsilon_{\bf k}}
\;\;\;\; , 
\label{eh.bis1}
\end{eqnarray}
\noindent
with the single quasiparticle dispersion relation $\epsilon_{\bf k} = \sqrt{ \xi_{\bf k}^2 + | \Delta_{\bf k}|^2}$. 
In the following, when we refer to Eqs.(\ref{eh.bis1}) when addressing the system dynamics, we keep $N$ finite. 
At variance, to recover the thermodynamics of the system, we refer to the 
large-$N$ limit of  Eqs.(\ref{eh.bis1}), in which they become the ``standard'' integral equations for the superconducting gaps within SCMF 
approximation, with $\frac{1}{N}\sum_{\bf k} \to \int_{\rm B.Z.} \frac{d^2k}{(2\pi)^2}$, with the integral taken 
over the full Brillouin zone.

At a given ${\bf k}$, the eigenvalues of $H_{\rm MF}$ corresponding to Bogoliubov quasiparticle excitations are
given by $ \pm \epsilon_{\bf k} \equiv \pm \sqrt{ \xi_{\bf k}^2+|\Delta_{\bf k}|^2}$, with the corresponding fermion operator
eigenmodes $\Gamma_{{\bf k},\pm}$ determined by  the Bogoliubov-Valatin transformation as 

 \beq
\left[ \begin{array}{c} \Gamma_{{\bf k},+}\\ \Gamma_{{\bf k},-} \end{array}\right] = 
\left[\begin{array}{cc} \cos \left(\frac{\theta_{\bf k}}{2}\right) & -e^{i\phi_{\bf k}}\sin \left(\frac{\theta_{\bf k}}{2}\right) \\
e^{- i\phi_{\bf k}}\sin \left(\frac{\theta_{\bf k}}{2}\right) & \cos \left(\frac{\theta_{\bf k}}{2}\right) 
\end{array}  \right] \left[\begin{array}{c} c_{{\bf k},\uparrow} \\  c_{-{\bf k},\downarrow}^\dagger   \end{array} \right]
\;\;\; , 
\label{sc.2a}
\eneq
\noindent
and  the parameters $\theta_{\bf k},\phi_{\bf k}$ defined by

\begin{eqnarray}
\xi_{\bf k} &=& \epsilon_{\bf k} \cos (\theta_{\bf k}) \nonumber \\
\Delta_{\bf k} &=& \epsilon_{\bf k} \sin (\theta_{\bf k}) e^{i\phi_{\bf k}}
\;\;\; . 
\label{sc.3a}
\end{eqnarray}
\noindent
 We now discuss the various superconducting phases that can set in on varying the parameters
 of  $H_{\rm MF}$ and the corresponding 
 phase diagram of the system. 

\subsection{Superconducting phases and phase diagram}
\label{supha}

In this section,  
 we derive the phase diagram of the system as a function of 
$U,V$, and $Z$, by holding $t'$ and $\mu$ fixed at selected value(s). To do so, we employ Eqs.(\ref{eh.bis1}) 
to determine $\Delta_S,\Delta_{x^2-y^2}$, and $\Delta_{xy}$     at 
a given value of the various system parameters.  

\begin{figure}
 \center
\includegraphics*[width=.9 \linewidth]{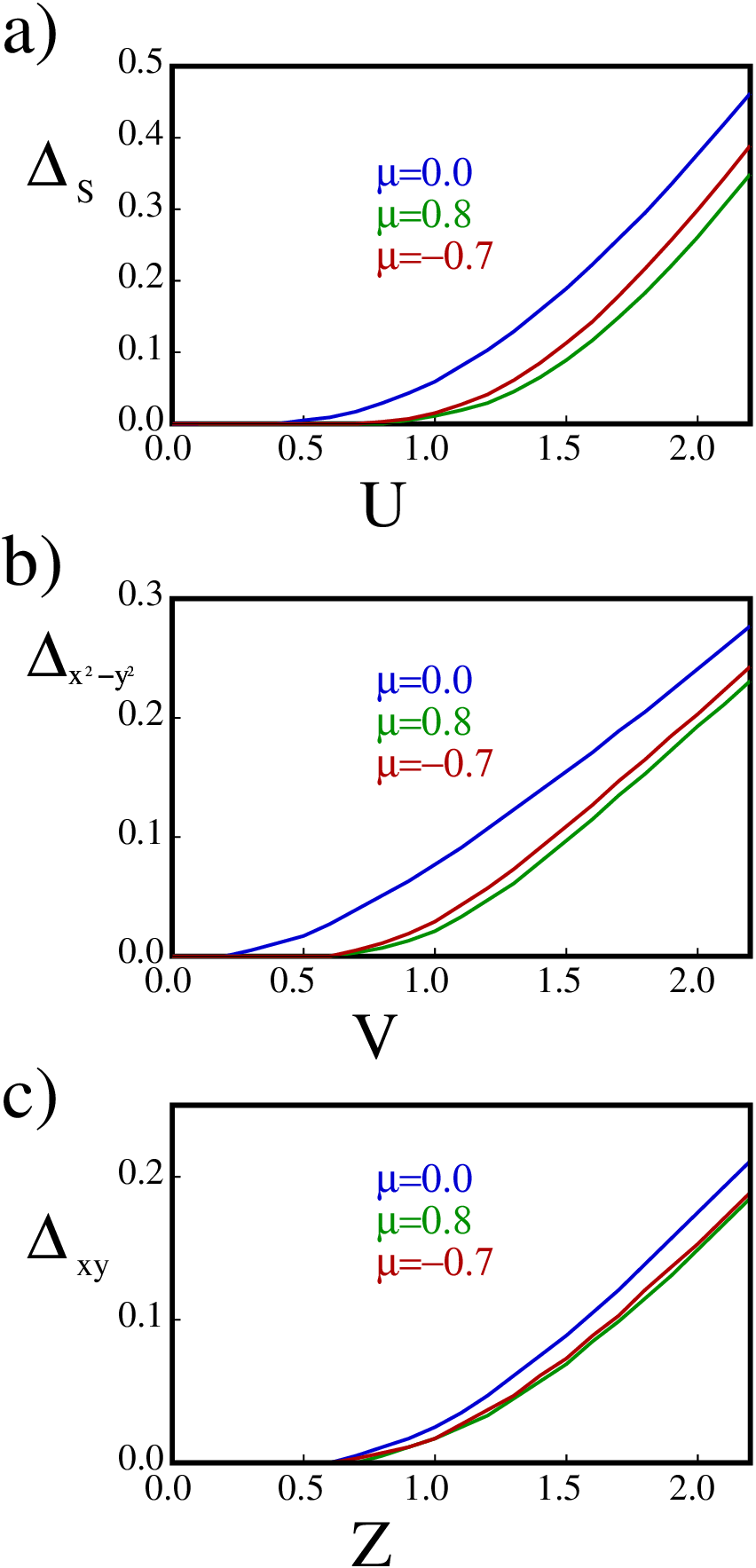}
\caption{ {\bf a):} $\Delta_S$ as a function of $U$ computed from Eqs.(\ref{eh.bis1}) by setting  $t'=V=Z=0$ and 
$\mu=0$ (blue line), $\mu=0.8$ (green line) and $\mu=-0.7$ (red line).
{\bf b):} $\Delta_{x^2-y^2}$  as a function of $V$ computed from Eqs.(\ref{eh.bis1}) by setting $t'=U=Z=0$ and 
$\mu=0$ (blue line), $\mu=0.8$ (green line) and $\mu=-0.7$ (red line). 
{\bf c):} $\Delta_{xy}$  as a function of $Z$ computed from Eqs.(\ref{eh.bis1}) by setting $t'=U=V=0$ and 
$\mu=0$ (blue line), $\mu=0.8$ (green line) and $\mu=-0.7$ (red line).
 }
\label{X_wave_gaps}
\end{figure}

In particular, we first of all study the phase diagram obtained by 
setting two of the three interaction strengths to 0 and increasing   the third one.
In this case, we always find a 
critical value of the variable interaction strength,
 beyond which the corresponding superconducting phase sets in.
We draw the corresponding phase diagrams   
 in Fig.\ref{X_wave_gaps}, where we plot 
$\Delta_S$ as a function of $U$,   for  $V=Z=0$ (panel {\bf a)}), 
$\Delta_{x^2-y^2}$ as a function of $V$,  for  $U=Z=0$ (panel {\bf b)}),
and $\Delta_{xy}$ as a function of $Z$ for $U=V=0$ (panel {\bf c)}),  for $\mu = 0,\mu=0.8$, and 
$\mu=-0.7$, respectively, with  $t'=0$.   In all three cases, we identify the superconducting 
phase transition, corresponding to the order parameter developing  a nonzero value as soon as the corresponding interaction strength 
 becomes greater than  a finite critical value. As a function of the chemical potential, the critical value is 
 recovered  by solving Eqs.(\ref{eh.bis1})  at a given $\mu$. In particular, from the plots drawn at different 
 values of $\mu$, we see how,  
as expected \cite{Micnas1990}, the tendency of the system to develop superconducting order is maximal at 
half-filling ($\mu =0$), while it gets lower as $\mu$ is moved to  either positive or negative values.

 As a next step, we now turn on  two different interactions strengths,  by holding at 
 zero the third one. In this case, it  is possible to recover (at least at SCMF level) phases with 
 two out of  $\Delta_S,\Delta_{x^2-y^2}$ and $\Delta_{xy}$ being $\neq 0$.
 The importance of phases as such has been, in fact,  argued to play  a crucial role  
 in the physics of high-temperature
 superconductors    \cite{Tsuei2000,Laughlin1998,Balatsky1998,Gorkov2001}. Moreover, 
 the two-gap coexistence 
 can lead to topologically nontrivial superconducting phases, such as the d+id superconductor \cite{Chern2016}. 
 Finally, as we discuss in the following, having (at least) two superconducting gaps $\neq 0$ is 
  an indispensable prerequisite to recover a
DPT between superconducting phases (including topologically nontrivial ones), 
along the time evolution of the nonequilibrium system  
\cite{Lepori2021,shorter_paper}. 
  
\begin{figure}
 \center
\includegraphics*[width=.9 \linewidth]{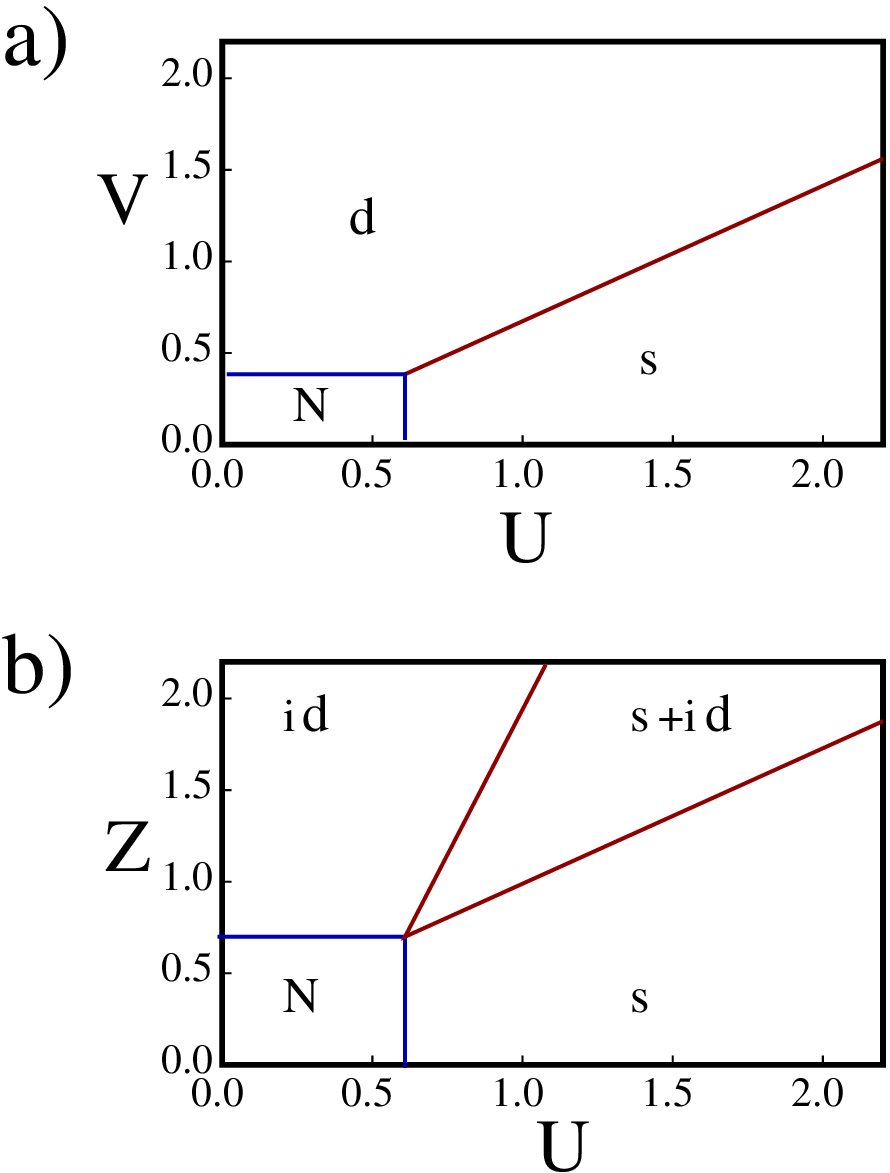}
\caption{ {\bf a):} Phase diagram in the $U-V$ plane computed from Eqs.(\ref{eh.bis1}) by setting   $\mu=t'=Z=0$.
{\bf b):} Phase diagram in the $U-Z$ plane  computed from Eqs.(\ref{eh.bis1}) by setting   $\mu=t'=V=0$. In the figure, 
N,s,d,id, and s+id respectively denote the normal phase (no superconducting gap), an s-wave superconducting phase 
(only $\Delta_S\neq0$), a  d-wave superconducting phase (only $\Delta_{x^2-y^2}\neq0$), an id-wave 
superconducting phase (only $\Delta_{xy}\neq0$), and the s+id-phase, with both $\Delta_S$ and $\Delta_{xy}$ $\neq0$.}
\label{phdiag}
\end{figure}

 As specific model calculations, in Fig.\ref{phdiag}{\bf a)} we show the phase diagram in the 
  $U-V$-plane at $Z=\mu=t'=0$. In this case, from Eqs.(\ref{eh.bis1}) we first 
  of all find a normal (N) phase  for $U<U_c$ and $V<V_c$, with (for $\mu=0$)
  $U_c \approx 0.6$ and $V_c\approx 0.35$, and $\Delta_S=\Delta_{x^2-y^2}=\Delta_{xy}=0$.
  On either increasing $U$ at fixed (and small)  $V$, or $V$ at fixed (and small) $U$, we respectively find 
  a purely s-wave superconducting phase with $\Delta_S \neq 0, \Delta_{s^2-y^2}=\Delta_{xy}=0$, and 
  a purely d-wave phase, with   $\Delta_{x^2-y^2}=0,\Delta_S=\Delta_{xy}=0$.  For large $U$ and $V$ 
  of comparable magnitude, we here find no phase where both  $\Delta_S$ and $\Delta_{x^2-y^2}$ 
  are $\neq 0$. In fact,  the system undergoes a direct phase transition from the s-wave to the d-wave 
  superconducting phase (or vice versa).  Of course, we note  that this is a specific result we obtained 
  within our SCMF approach. While it is unlikely that a better estimate of the effects of the  fluctuations
 might stabilize a mixed s+d phase, yet, pertinent modifications of our model Hamiltonian (which go beyond the scope
 of our paper), including additional  hoppings and/or interactions, would likely stabilize it.  
  
At variance, as  we show in Fig.\ref{phdiag}({\bf b)}, for $V=0$, we find, for $\mu=0$, 
 $U_c \approx 0.6$ and $Z_c\approx 0.7$.
However, in this case, when both $U$ and $Z$  are $\neq 0$ and $V=0$, in addition to the ``pure'' $s$-wave and $id$-wave phases, we 
do find a coexistence phase with both $\Delta_S$ and $\Delta_{xy}$ $\neq 0$ ($s+id$-phase). This is also what happens 
when  $U=0$ and both $V$ and $Z$ are $\neq 0$, where the corresponding $d+id$-phase also exhibits 
nontrivial topological properties \cite{shorter_paper}. At $\mu \neq 0$ one finds that, 
consistently with the results reported in Fig.\ref{X_wave_gaps}, the  nonzero chemical potential  
just   determines a mild shrinking of the superconducting regions:
a feature that  does not substantially affect the main qualitative 
aspects  of the phase diagrams reported in Fig.\ref{phdiag}.

 Finally, we point out that, although, for $V=Z=0$ and at half-filling, the superconducting state
is degenerate in energy with a charge density wave phase, as soon as a nonzero negative $V$ and/or $Z$ is 
turned on and/or the system is tuned out of half-filling ($\mu \neq 0$), the superconducting phase comes out
to be always more stable than the charge density wave one \cite{Micnas1990}. Consistently with the above conclusion, 
throughout our paper we focused onto superconducting phases only, although with different possible 
kinds of gap order parameter.

Given a phase diagram such as the ones we show in Fig. \ref{phdiag}, a protocol leading to 
 a DPT  can, in principle, be realized by simply preparing the system in an initial state 
within a given phase and by quenching, at $t=0^+$,  the interaction parameters to a point
within a different phase in the phase diagram.

  As we evidence above, 
 the real-time evolution induces an effective dependence on time in the superconducting gap order parameter 
 \cite{Peronaci2015}.    The time-dependent superconducting gap can be tuned and possibly observed 
 in, e.g., an out-of-equilibrium pump-probe experiment. In such an experiment, the pump pulse induces a change in the  gap. At
 the same time,  the reflectivity and the optical conductivity can be measured with a second probe pulse at different 
 pump-probe time delays. The saturated reflectivity and the gap in the real part of the optical conductivity make it possible to monitor
 the magnitude of the superconducting gap as a function of time  \cite{Mitrano2016,Nava2018}. It is also possible to experimentally 
 adjust the interaction strengths $U, V$, and $Z$, by tuning the electron-phonon coupling like, for example, in synthetic crystals \cite{Choi2023},
  or in time and angle resolved photoemission spectroscopy experiments \cite{Jianwei2023}.

\subsection{Lindblad master equation}
\label{lineq}

 We now review the LME approach, which we employ    to describe the 
dynamics of the nonequilibrium open  system. 

As stated above, our protocol 
for inducing the relaxation dynamics in the nonequilibrium system consists
first in quenching, at $t=0$, the interaction strengths from their initial values 
$U^{(0)},V^{(0)},Z^{(0)}$ (not necessarily all $\neq 0$), from which we determine  
the initial state of the system, to $U^{(1)},V^{(1)},Z^{(1)}$. Along the 
derivation of Ref.\cite{Peronaci2015}, we study the dynamics of our system within 
 a time dependent version  of the SCMF approximation, based on the LME approach.  This approach
recovers the dissipative dynamics   induced in the system by the interactions between quasiparticles beyond BCS 
approximation, and/or by the coupling between the fluctuations of the order parameter and the quasiparticle continuum 
\cite{Yuzbashyan2005,Yuzbashyan2006,Yuzbashyan2006b,Cui2019}. Following  Refs. \cite{Nava2021,Nava2023} and  
using   $H_{\rm MF}$ in Eq.(\ref{eh.2a}) as our main system Hamiltonian, we  
write down  the full set of  LME for the time evolution of the density matrix operator 
of the system coupled to the bath, $\rho (t)$, which we pertinently complement by self-consistently 
recalculating, at any $t$,  the (time-dependent) superconducting order parameter $\Delta_{\bf k} (t)$.
  Eventually, we show that our systematic approach 
is perfectly consistent with the one introduced in Ref.\cite{Cui2019} on phenomenological grounds.

The LME for  $\rho (t)$ has the form    

\begin{widetext}
\begin{eqnarray}
\frac{d \rho (t)}{dt}&=&-i[H_{\rm MF} (t),\rho(t)]+g\sum_{\lambda = \pm}\sum_{\bf k} (1-f(\lambda \epsilon_{\bf k} (t) ))\{ [ \Gamma_{{\bf k},\lambda} (t) ,\rho (t)\Gamma_{{\bf k},\lambda}^\dagger (t) ]
-[\Gamma_{{\bf k},\lambda}^\dagger  (t) , \Gamma_{{\bf k},\lambda} (t) \rho (t)]\} \nonumber \\
&+& g \sum_{\lambda=\pm}\sum_{\bf k} f(\lambda \epsilon_{\bf k} (t) ) \{ [\Gamma_{{\bf k},\lambda}^\dagger (t),\rho (t) \Gamma_{{\bf k},\lambda}(t) ]-[\Gamma_{{\bf k},\lambda} (t),\Gamma_{{\bf k},\lambda}^\dagger(t) \rho (t)]\} 
\:\:\:\: .
\label{leq.1}
\end{eqnarray}
\noindent
\end{widetext}
In Eq.(\ref{leq.1}) we have denoted with $g$ the strength of the coupling between the system and the external bath. Moreover, 
we have set the coupling strength corresponding to the quasiparticle annihilation and creation 
operators, $\Gamma_{{\bf k},\lambda}$ and $\Gamma_{{\bf k},\lambda}^{\dagger}$ (see Eq.(\ref{sc.2a})), so
to make them proportional to $(1-f(\lambda \epsilon_{\bf k}))$ and 
$f(\lambda \epsilon_{\bf k})$, respectively,  with $f(\epsilon )$
  being the Fermi distribution function.
  Accordingly, Eq.(\ref{leq.1}) describes the system coupled to a bath with which it can exchange both energy and matter, through the injection or the annihilation of Bogoliubov quasiparticles. Indeed, Lindblad jump operators describe the creation/annihilation of  these quasiparticles with, as stated above,  a transition probability chosen so to  satisfy the detailed balance condition and to make 
   the stationary state of the LME  to be described by a thermal grandcanonical density matrix. Our choice is a particular case of 
 the generic system-bath Hamiltonian [as shown in Eq. (3.128) of Ref. \cite{Petruccione2002}], which is realized as 
 a pertinent  linear combination of the tensor products between  system and bath eigenstates 
 [note that changing the linear combination would only affect   the numerical values of the coupling strengths, not
 the general form, of   Eq. (\ref{leq.1})]. 
    
   While, in principle, we could arbitrarily choose
 the Lindblad jump operators and the corresponding
  coupling strengths, setting them as we do here, we make sure that the detailed balance is ensured and the Boltzmann distribution 
 is a stationary solution of the Lindblad equation \cite{Petruccione2002,Nava2019}.  Moreover, as we discuss below, 
 our choice eventually yields results for the time evolution and for the asymptotic alternative states of our system that are
 perfectly consistent with the phenomenological approach of Ref. \cite{Cui2019}.   
Since we self-consistently compute the 
 superconducting order parameter, $\Delta_{\bf k}(t)$, at any time $t$,  $H_{\rm MF} (t)$ at the right-hand side of Eq.(\ref{leq.1}) 
 acquires an explicit dependence on $t$ and, accordingly, its eigenvalues [$\pm \epsilon_{\bf k}(t)$] and the corresponding 
 eigenmodes [$\Gamma_{{\bf k},\pm }(t)$] depend on $t$, as well. 
 
 To write the SCMF equation for $\Delta_{\bf k} (t)$,  we take 
  advantage of the fact that $H_{\rm MF} (t)$ is quadratic in the quasiparticle operators and that the coupling to the external bath is linear
 in the same operators. This allows us to   employ Eq. (\ref{leq.1}) to write a closed set of equations for the 
 (time-dependent) average values of  the products of two single-fermion operators. Specifically, we set 
  
 \begin{eqnarray}
 \nu_{{\bf k},\sigma} (t) &=& \sigma {\rm Tr} \left[ \rho (t)  \left( c_{{\bf k},\sigma}^\dagger c_{{\bf k},\sigma} -\frac{1}{2} \right) \right] \: , \nonumber \\
f_{{\bf k} }(t)&=& {\rm Tr} [\rho (t) c_{{\bf k},\downarrow} c_{{\bf - k},\uparrow} ] 
\;\;\; .
\label{leq.2}
\end{eqnarray}
\noindent
 We now point out that, on one hand,   there is zero spin polarization in the initial state, on the other
hand,   no spin polarization can either be 
generated along the dynamical evolution of the system, as described by Eq.(\ref{leq.1}). Indeed, this can be 
readily verified by introducing the total spin operator ${\bf S}=\sum_{\bf k} {\bf S}_{\bf k}$, with 
 the Anderson isospin operator at given ${\bf k}$, ${\bf S}_{\bf k}$,  defined as. 

\beq
S^a_{\bf k} = \frac{1}{2} [c_{{\bf k},\uparrow}^\dagger , c_{{\bf -k},\downarrow}] \sigma^a \left[\begin{array}{c} c_{{\bf k},\uparrow}\\
c_{{\bf -k},\downarrow}^\dagger \end{array}\right]
\;\; , 
\label{exspin.1}
\eneq
\noindent
 $\{ \sigma^a \}$ being the Pauli matrices. At time $t$, we obtain $\langle {\bf S} (t)\rangle ={\rm Tr} [\rho (t) {\bf S}]$. 
 From Eq.(\ref{leq.1}), taking into account that $[H_{\rm MF} (t) ,{\bf S}]=0$ and that the quasiparticle operators 
 $\Gamma_{{\bf k},\lambda}(t)$ carry a well-defined spin content, it can be readily shown that 
 $\frac{d \langle {\bf S} (t) \rangle}{dt}=0$, which implies  
 $\nu_{{\bf k},\uparrow} (t)=-\nu_{{\bf k},\downarrow}(t)\equiv \nu_{\bf k}(t)$. As a result, 
we recover, in the zero-temperature limit,  the (closed) set of differential equations 

\begin{eqnarray}
\frac{ d \nu_{{\bf k} } (t)}{dt}&=& -\frac{g \xi_{\bf k}}{\epsilon_{\bf k}(t)} -2 g\nu_{\bf k}(t) + \Im m\{ [    \Delta_{\bf k} (t) ]^*   f_{\bf k} ( t) \}  
\label{leq.3} \\
\frac{ d f_{\bf k} (t)}{dt}&=& -(2i \xi_{\bf k} +2g) f_{\bf k}(t) - 2 i \Delta_{\bf k} (t) \nu_{\bf k}(t) + \frac{g \Delta_{\bf k} (t)}{\epsilon_{\bf k}(t)} 
  \:\nonumber ,
\end{eqnarray}
\noindent
with $\epsilon_{\bf k}(t)=\sqrt{\xi_{\bf k}^2 + |\Delta_{\bf k}(t)|^2}$ and $\Im m$ denoting the imaginary part.  To compute 
$\Delta_{\bf k} (t)$ we resort to the time-dependent SCMF approach. 
 This corresponds to a time-dependent generalization of the BCS variational ansatz,
which is equivalent to assuming a time dependent generalization of   the latter one of Eqs.(\ref{eh.4a}) in the form

\begin{eqnarray}
 \Delta_{\bf k} (t) &=& \Delta_S (t)  + 2 \Delta_{x^2-y^2}(t) \{\cos(k_x) - \cos (k_y) \} 
\nonumber \\
 &-& 4 i \Delta_{xy} (t) \sin (k_x ) \sin (k_y ) 
\:\:\: . 
\label{leq.4b}
\end{eqnarray}
\noindent
The parameters $\Delta_S (t),\Delta_{x^2-y^2}(t)$, and $\Delta_{xy}(t)$ have to be self-consistently computed 
by employing a pertinent, time-dependent, generalization of Eqs.(\ref{eh.bis1}) by replacing $\Delta_{\bf k}/\epsilon_{\bf k}$ at
the right-hand side of the equations with $f_{\bf k} (t)$ obtained by solving Eqs.(\ref{leq.3}). 

 To further ground 
the time-dependent SCMF approach leading to Eq.(\ref{leq.4b}), we note that the same results as 
the ones recovered within our approach were derived   in Ref.\cite{Peronaci2015} within Keldysh nonequilibrium 
approach, in the limit of a small change in the interaction strengths.  

As we pointed out above,  differently from the derivation of Ref.\cite{Peronaci2015},
 in our approach, the direct 
coupling to the external bath always determines a finite relaxation timescale ($\sim (2g)^{-1}$) for the superconducting order parameter.
This uniquely sets the  asymptotic value of $\Delta_{\bf k} (t)$ as $t \to \infty$ to the one corresponding to the equilibrium 
superconducting phase described by $H$ in Eq.(\ref{eh.1a}) with   interaction strengths $U^{(1)},V^{(1)},Z^{(1)}$. 
As we discuss in the following, when taking the system across a DPT, 
the coupling to the external bath is also crucial in setting the time  $t_*$ at which the transition takes place.

In order to physically ground our choice for the Lindblad operators entering the LME in Eq.(\ref{leq.1}), we 
now compare our formalism with the phenomenological approach of Ref.\cite{Cui2019} (to which we refer for 
a systematic discussion about the relation between the terms of the phenomenological equation -- and, therefore, 
of the LME -- and the microscopic quasiparticle dynamics). To do so, we 
employ Eqs.(\ref{leq.3}) (which are a direct consequence 
of the LME in Eq.(\ref{leq.1})) we can, therefore, write down the equations of motion for ${\bf S}_{\bf k} (t)$ in 
Eq.(\ref{exspin.1}) as 

\beq
\frac{ d \langle {\bf S}_{\bf k} (t) \rangle}{ d t } = \langle {\bf S}_{\bf k} (t) \rangle \times {\bf B}_{\bf k} (t)  - 2 g \langle {\bf S}_{\bf k} (t) \rangle 
+ 2 g \langle {\bf S}_{{\bf k},*} (t) \rangle 
\;\;\;\; , 
\label{ccc.1}
\eneq
\noindent
with 

\beq
{\bf B}_{\bf k} (t) = \left[ \begin{array}{c} \Re e [-\Delta_{\bf k} (t ) ] \\ \Im m [ \Delta_{{\bf k}} (t ) ] \\ \xi_{\bf k} \end{array} \right]
\;\;\;\; , 
\label{ccc.2}
\eneq
\noindent
and 

\beq
 \langle {\bf S}_{{\bf k},*} (t) \rangle = \frac{1}{2 \epsilon_{\bf k} (t)} {\bf B}_{\bf k} (t) 
 \:\:\:\: . 
 \label{ccc.3}
 \eneq
 \noindent
 From Eq.(\ref{ccc.3}) we infer that the vector $\langle {\bf S}_{{\bf k},*} (t) \rangle$ is always proportional 
 to ${\bf B}_{\bf k}(t)$, that is, fully longitudinal. Thus, we conclude that 
 Eq.(\ref{ccc.2}) has exactly the same form as Eq.(9) of Ref.\cite{Cui2019}, provided, in the formalism of that 
 paper, one takes the longitudinal ($T_1^{-1}$) and transverse ($T_2^{-1}$) relaxation rates for 
 ${\bf S}_{\bf k}$ according to $T_1^{-1}=T_2^{-1}=2g$. In fact, finite values of $T_1^{-1}$ and $T_2^{-1}$ in a 
 nonequilibrium superconductor have been argued to be related to the (inverse) timescales of integrability-breaking
 (that is, non BCS-like) interactions. Specifically, $T_1$ is related to the interaction among quasiparticles, while 
 $T_2$ to the direct coupling between the quasiparticles and the fluctuations of the superconducting order parameter \cite{Cui2019}. 
 In general, both $T_1$ and $T_2$ must be regarded as phenomenological parameters, and their  values depend 
 on the specific material and on the protocol implemented in the measurement. For instance, in the case in 
 which nonequilibrium is induced by acting with strong optical pulses with Terahertz frequencies on NbN, or on 
 Nb$_3$Sn, typical values of the order of 10 ps have been fitted from the experiments discussed in 
  Ref.\cite{Cui2019}, with a pulse duration of a few ps. Assuming, in our model, an over-all energy scale 
  $J\sim 1 {\rm eV}$ would yield $g \sim 0.002$. However, since, within our protocol, we assume
  that the superconductor is adiabatically  prepared in the nonequilibrium state, starting from $t\to -\infty$, we may
  expect, in a realistic system, values of $g$ that are significantly  larger than the previous
  estimate. Consistently with the uncertainty on its actual value in a realistic system, we perform our calculations 
  for at least two values of $g$, typically different by   orders of magnitude from each other. 

  In both cases the bath is a gas of Bogoliubov quasiparticles. In the self-consistent time evolution, the bath is intrinsic 
to the system and the LME accounts for residual interactions between the Bogoliubov quasiparticles neglected in the mean-field 
BCS approach \cite{Cui2019}; in the non self-consistent time evolution the proximity effect may allow, 
for instance, for quasiparticles to be exchanged  between 
the system and an underneath superconductor at equilibrium \cite{Efetov2008}.  

In the following, we present and discuss our results for the time evolution of the superconducting order parameter in the system coupled 
to the external bath in some   paradigmatic cases. Also,  in Appendix \ref{sudden} we review the same calculation
for the case in which, at $t=0$, one directly quenches $\Delta_{\bf k} (t)$. Besides being useful for comparison with 
the case in which one quenches the interaction strengths, this latter approach is of great relevance in our
 calculation of   the spin-Hall conductance in Ref.\cite{shorter_paper}. 
   
\section{Time evolution of the superconductor coupled to the external bath}
\label{timevol}

We now discuss the time evolution of our nonequilibrium  open system.
  Specifically,  we initialize the system in the groundstate 
of $H_{\rm MF}$ with an assigned value of the gap parameter $\Delta_{\bf k}^{(0)}$, corresponding to 
the state realized at different values of the interaction strengths, $U^{(0)}, V^{(0)}$, and $Z^{(0)}$. 
Then,   at $t=0^+$,  we quench the interaction strengths to $U^{(1)}, V^{(1)}$, and $Z^{(1)}$ and, at the same time,  
 we turn on the coupling $g$ to the  bath. For $t>0$ the system  evolves
toward its asymptotic state, and the superconducting gaps explicitly depend on $t$  according to 
Eqs.(\ref{leq.4b}).  
 
  Along our analysis, we first consider the case in which only one of the three interaction 
  strengths is $\neq 0$ and, at a second stage, we generalize our derivation to the case in which 
two interaction strengths become $\neq 0$.  This eventually allows us to 
investigate whether a DPT is expected to set in along the time evolution of the system 
and,  if so, what are its main features.

Throughout our derivation we work in the zero-temperature limit.  In this limit, 
the function $f(\lambda \epsilon_{\bf k} (t) )$ in the coupling strengths in front of the Lindblad operators in
 Eq.(\ref{leq.1}) is either equal to 0, if $\lambda=+1$, or to $1$, if $\lambda=-1$, regardless 
 of $t$. While  this provides a 
remarkable simplification of our derivation below, yet, following our above
analysis, it is in principle straightforward to address the finite temperature case as well.  
 
\subsection{Relaxation dynamics of a single-component order parameter}
\label{relaxingle}

We begin by keeping only one among the interaction strengths $U,V$ and $Z$ to be $\neq 0$.

\begin{figure}
 \center
\includegraphics*[width=.9 \linewidth]{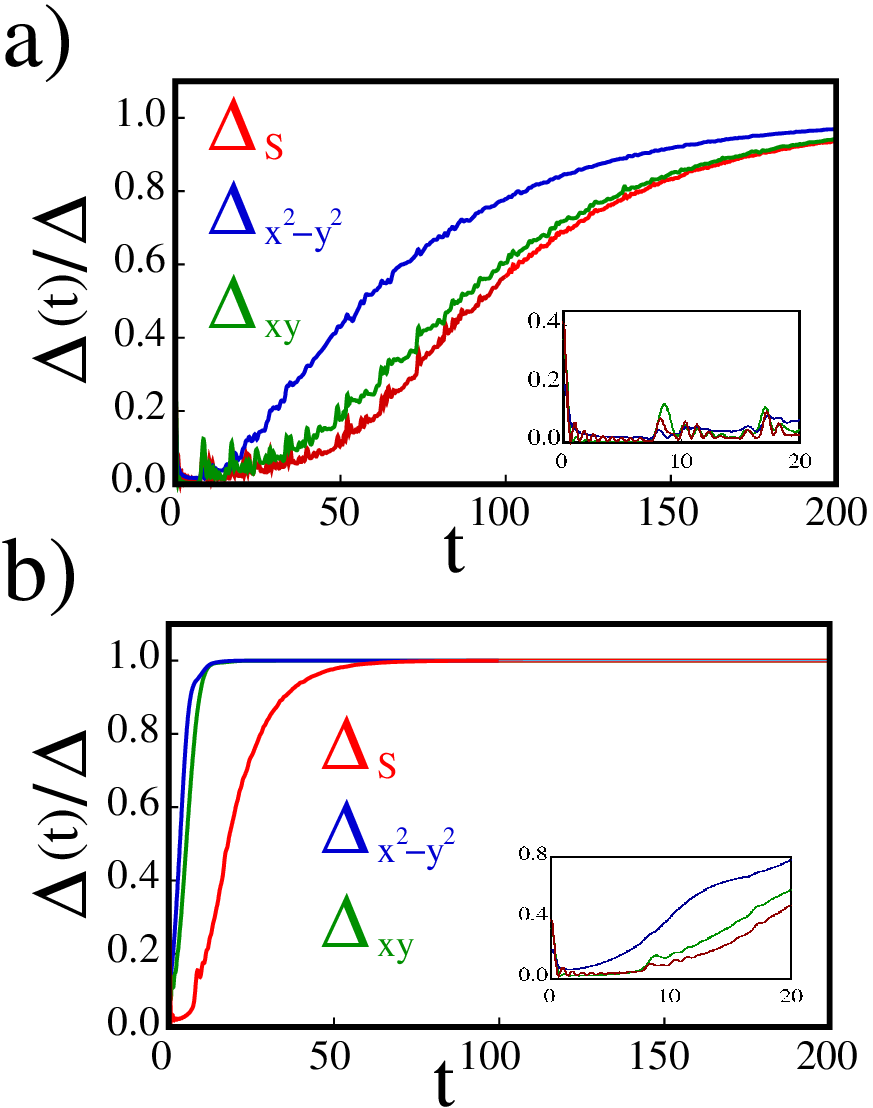}
\caption{ {\bf a):} Time evolution of the rescaled order parameters $\Delta (t )/ \Delta $ self-consistently computed for 
  $g=0.01$, for the case  in which $\Delta_S (t=0) =0.0750$ and $\Delta_{x^2-y^2}(t=0)=\Delta_{xy}(t=0)=0$
(computed at $U=1.5$, $V=Z=0$ -- red curves),  for the case in which $\Delta_{x^2-y^2}  (t=0) =0.0607$ and
 $\Delta_{S}(t=0)=\Delta_{xy}(t=0)=0$
(computed at $V=1.5$, $U=Z=0$ --blue curves), and for the case in which $\Delta_{xy} (t=0) =0.1208$ and
 $\Delta_{x^2-y^2}(t=0)=\Delta_{S}(t=0)=0$ (computed at $Z=1.5$, $U=V=0$ -- green curves)
 [Inset: zoom of the plots restricted to the interval $0\leq t \leq 20$]. 
 {\bf b):} Same as in panel {\bf a)}, but with $g=0.05$.}
\label{single_gap_sc}
\end{figure}

\begin{figure}
 \center
\includegraphics*[width=.9 \linewidth]{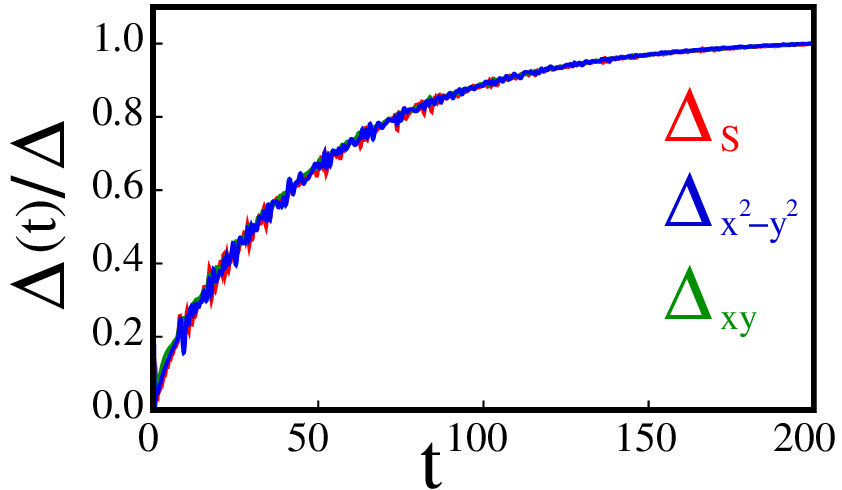}
\caption{ Time evolution of the rescaled order parameters $\Delta (t)/\Delta $ non self-consistently computed for 
 $g=0.01$, for the case  in which $\Delta_S (t=0) =0.075$ and $\Delta_{x^2-y^2}(t)=\Delta_{xy}(t)=0$
(computed at $U=1.5$, $V=Z=0$ -- red curve),  for  the case in which $\Delta_{x^2-y^2}  (t=0) =0.030$ and $\Delta_{S}(t)=\Delta_{xy}(t)=0$
(computed at $V=1.5$, $U=Z=0$ --blue curve), and for  the case in which $\Delta_{xy} (t=0) =0.0302$ and $\Delta_{x^2-y^2}(t)=\Delta_{S}(t)=0$
 (computed at $Z=1.5$, $U=V=0$ -- green curve).  }
\label{quench_single}
\end{figure}

In Fig.\ref{single_gap_sc} we plot the superconducting gap, normalized to  its asymptotic (that is,
$t\to\infty$) value, for the case in which $\Delta_S (t) \neq 0$ and $\Delta_{x^2-y^2}(t)=\Delta_{xy}(t)=0$
(red curves),  in which $\Delta_{x^2-y^2}  (t) \neq 0$ and $\Delta_{S}(t)=\Delta_{xy}(t)=0$
 (blue curves), and for the case in which $\Delta_{xy} (t) \neq 0$ and $\Delta_{x^2-y^2}(t)=\Delta_{S}(t)=0$ (green curves). 
We respectively set     $g=0.01$ (Fig.\ref{single_gap_sc}{\bf a)}), and 
$g=0.05$  (Fig.\ref{single_gap_sc}{\bf b)}).  Here, as basically anywhere else below, we 
set  $t'=\mu=0$.  From Fig.\ref{single_gap_sc}, we see that, 
for any one of the three gaps, the relaxation rate
is   solely determined by the coupling to the bath: the larger is $g$, the 
faster is the relaxation of the superconducting order parameter toward its asymptotic value. 
In addition, we also note a remarkable dependence of the relaxation time on the 
symmetry of the order parameter. This is demonstrated by the different shape of the curves
for different gaps, which is  apparent  in both cases,
although it is much more evident in Fig.\ref{single_gap_sc}{\bf a)} due to the smaller value of $g$ and to 
the correspondingly slower relaxation of the superconducting gaps.  Remarkably, a similar effect  
also appears for a closed system ($g=0$) \cite{Peronaci2015}. 
It is likely related to different dissipation mechanisms
  that  set in along the relaxation of the order parameter. Such effects are, in general, 
 well-captured by the time-dependent SCMF approach. At variance, if one gives up self-consistency and 
 simply ``quenches'' the superconducting gap at $t=0$  (see Appendix \ref{sudden} for details),  any dependence
 on the symmetry of the superconducting order parameter is washed out. 
To evidence this point, in   Fig.\ref{quench_single}, we draw plots similar to the ones in
Fig.\ref{single_gap_sc} but by giving up self-consistency. Indeed, we then see no appreciable dependence of 
  the time dependent superconducting order parameter on its symmetry. 

Another remarkable feature of the time evolution of $\Delta_{\bf k} (t)$ is given by the oscillations in the amplitude 
of the superconducting order parameter. While they have been already noticed and discussed in 
Ref.\cite{Peronaci2015}, in our specific case they exhibit a peculiar behavior, due to the nonzero coupling 
to the  bath.  As the system is prepared in a nonequilibrium state that, in principle, has a nonzero overlap with all the
excited states of the Hamiltonian that determines the time evolution for $t>0$, we expect, for small time 
intervals, oscillations in the amplitude of the order parameters over several frequencies. To evidence that 
this is, in fact, the case, in the inset of Fig.\ref{single_gap_sc}{\bf a)} we show the same plot as in the main 
figure, but restricted to the interval $0\leq t \leq 20$. We clearly see the expected oscillations which, 
as $t$ gets large, start to be damped by the finite value of $g$. A similar effect can be identified in the inset 
of Fig.\ref{single_gap_sc}{\bf b)}, although now the damping is much faster, due to the larger value of $g$.

\noindent

\noindent
      
\subsection{Relaxation dynamics of a two-component order parameter}
\label{rtwo}

 We now consider the case in which (at least) two interaction strengths are $\neq 0$. 
  
We now consider   the relaxation dynamics of a  system  prepared in the ground state of 
  $H_{\rm MF}$ in Eq.(\ref{eh.2a}), with $\Delta_S^{(0)} = \Delta_{x^2-y^2}^{(0)} =0$, 
and $\Delta_{xy}^{(0)}=0.03$, which corresponds  to having
  $U^{(0)} = V^{(0)} = 0$, $Z^{(0)} >0$. Moving across $t=0$, we quench the interaction strengths 
to $ (U^{(1)},  V^{(1)}, Z^{(1)} )= (1.5,0,1.5)$. As a result, the system develops a nonzero 
$\Delta_S (t)$ and $\Delta_{xy}(t)$, which we compute for two different values of $g$  and 
for $t'=\mu=0$. 
 
\begin{figure}
 \center
\includegraphics*[width=.9 \linewidth]{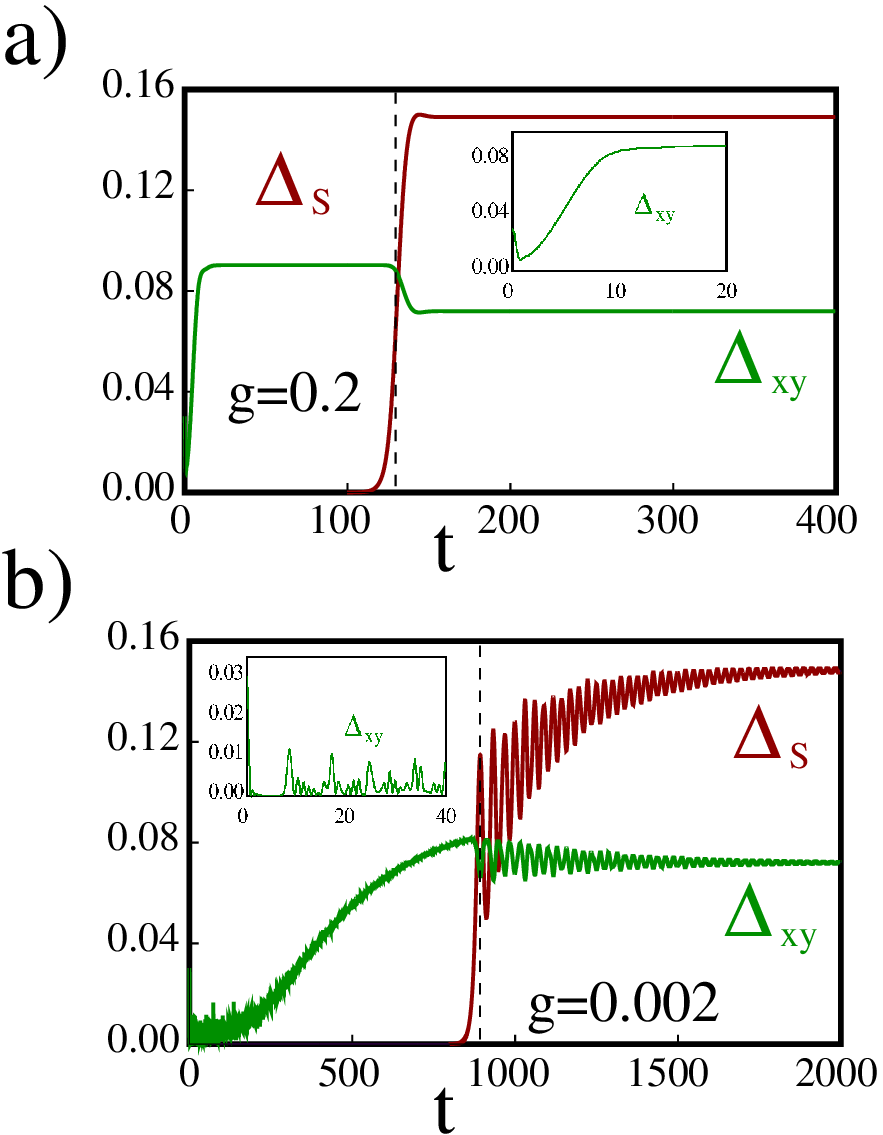}
\caption{{\bf a):} Time evolution of $\Delta_S (t)$ (red curve) and of $\Delta_{xy} (t)$ (green curve) computed 
in a system with   $U=Z=1.5$, $V=0$, coupled to a bath with interaction 
strength $g=0.2$ and prepared, at $t=0$, in a state with  $\Delta_{xy}^{(0)} \approx 0.03$ [{\bf Inset:}
The same plot (for $\Delta_{xy}(t)$ only), restricted to $0\leq t <20$].
 {\bf b)}: Same as in panel {\bf a)} but with $g=0.002$. In both cases the vertical  dashed 
lines mark  the onset of the DPT [{\bf Inset:}
The same plot (for $\Delta_{xy}(t)$ only), restricted to $0\leq t <40$].}
\label{sid_f}
\end{figure}

 In Fig.\ref{sid_f} we plot $\Delta_S (t)$ and $\Delta_{xy} (t)$. To evidence the effects of the coupling to the  
 bath on the time evolution of the superconducting gap, we perform the calculation for 
 $g=0.2$ (Fig.\ref{sid_f}{\bf a)}) and for $g=0.002$ (Fig.\ref{sid_f}{\bf b)}). We see that, on 
 one hand, there is, for the larger values of $g$, a suppression of the oscillations in the 
 superconducting gap. However,  in both cases we identify    a 
   finite interval of time $[0,t_*]$ within which $\Delta_S (t)$  
remains pinned at 0 and $\Delta_{xy} (t)$ keeps finite and  basically constant at large $g$, while it 
smoothly increases, with a fast oscillating modulation, at small $g$. Also, 
we note how  the ``critical time'' $t_*$ increases upon lowering   $g$. 
As $t$ goes across $t_*$,  $\Delta_S (t)$ jumps to a finite value. For $t>t_*$, for $g=0.2$, $\Delta_S (t)$ has a finite value, roughly constant.
  For $g=0.002$,  $\Delta_S (t)$ displays damped oscillations. 
 In both cases, however, we clearly see how, as $t\to\infty$, $\Delta_S (t)$ converges toward
the value $\Delta_{S,\infty} =0.15$. A similar trend is shown by $\Delta_{xy}(t)$, for $t>t_*$ which also
asymptotically flows to  $\Delta_{xy,\infty}=0.073$. Remarkably, $(\Delta_S , \Delta_{xy} ) =
 (\Delta_{S,\infty},\Delta_{xy,\infty})=(0.15,0.073)$
is exactly the set of values of the superconducting gap that one finds  from in the phase diagram of 
Fig.\ref{phdiag}{\bf b)} for $U=Z=1.5$. Thus, we conclude that the net effect of 
coupling the system to the bath is to trigger a time evolution of the superconductor between
two equilibrium phases,  an initial phase with $\Delta_S^{(0)}=\Delta_{x^2-y^2}^{(0)}=0$, $\Delta_{xy}^{(0)}=0.03$, 
and a final (asymptotic) phase with $\Delta_{S,\infty}=0.15$, $\Delta_{x^2-y^2}^{(1)}=0$, and 
$\Delta_{xy,\infty}=0.073$. Therefore, as a matter of fact, both plots in Fig.\ref{sid_f}
evidence a DPT in our system, whose precise location ($t=t_*$) does  depend on the 
value of $g$. In the following, we further corroborate our conclusion by studying the time dependence 
of the fidelity between the initial state of the system and the state that, at time $t$, is described by 
the density matrix $\rho (t)$ \cite{Zvyagin2016,Heyl2018,Heyl2019}.

 To summarize, from the time dependence of the superconducting 
 order parameters, we clearly find evidence for DPTs, basically determined by the mismatch between the 
 initial and the final state of the system. 
 To better ground our conclusions, in the following we estimate the fidelity ${\cal F}(t)$ along the time evolution, 
 finding an excellent consistency with  the conclusions 
 about the DPT we recovered from the time-dependent superconducting order parameters.
     
\section{Fidelity across   the dynamical phase transition}
\label{toprt}

In Section \ref{rtwo}  we inferred the emergence of the DPT  from 
the time dependence of the superconducting order parameter after quenching the 
interaction strengths.  In general,  in a closed nonequilibrium system that, 
at time $t$, is described by a pure state $|\psi(t)\rangle$, the
standard mean to analyze a DPT is   looking at nonanalyticities in 
 the Loschmidt echo ${\cal L} (t) = | \langle \psi ( 0 ) | \psi (t ) \rangle|^2$, with $|\psi(0)\rangle$
 being the initial state of the system  \cite{Zvyagin2016,Heyl2018,Heyl2019}.   In our case, for $t>0$, 
 the state  is described by the density matrix $\rho (t)$ which, in general,
 does not correspond to a pure quantum state. For this reason, 
we now characterize the DPT  by looking at nonalyticities in  
  the fidelity  ${\cal F} (t)$ between $|\psi (0)\rangle$ and  
  density matrix $\rho (t)$    \cite{Zvyagin2016,Heyl2018,Heyl2019,Wu2022}.  
  Specifically,  in our case ${\cal F} (t)$ is defined as \cite{Mera2018}

\beq
{\cal F} (t) = \langle \psi ( 0) | \rho (t) | \psi ( 0 ) \rangle
\;\;\;\; . 
\label{lale.1}
\eneq
\noindent
The time evolution of $\rho (t)$ for $t>0$ is determined according to the LME in Eq.(\ref{leq.1}). Due   to  the 
time-dependent self-consistency, Eq.(\ref{leq.1}) is effectively nonlinear and, therefore, it is quite a formidable 
task to solve it in practice, even for small lattices. For this reason, 
in the following we resort to a sequence of reasonable approximations, which eventually allow us to recast ${\cal F} (t)$ in 
a tractable form.  

To begin with, let us introduce the basis of the ${\cal N}$-particle many-body states created by the quasiparticle creation operators
determined by $H_{\rm MF} (t)$. Specifically, we set 

 \beq
 | {\cal N} ,  t \rangle_{  \{ {\bf q}_j \} , \{ \lambda_j \}  }  = \prod_{j=1}^{{\cal N}} [ \Gamma_{{\bf q}_j , \lambda_j } ( t ) ]^\dagger  | {\bf 0}  \rangle 
 \: ,
 \label{smalln.2}
 \eneq
 \noindent
 with the vacuum $|{\bf 0}\rangle$  defined by the condition  $ \Gamma_{{\bf q} , \lambda  } ( t )  | {\bf 0}\rangle  
 = 0$, $\forall {\bf q},\lambda$.  
  Now, on numerically integrating Eqs.(\ref{leq.3}) for $\nu_{\bf k} (t)$, we easily verify that, in the half-filled 
 system,  $\nu_{\bf k}(t)=0$ constantly,  
 along the time evolution.  Therefore, consistently with the result that, on average, 
 we get ${\cal N}=N$, we make the assumption that all the density matrix elements involving states
 with total filling different from $1/2$ are negligible and, then,  can be safely put equal to 0. This allows us to simplify  the right-hand 
 side of Eq.(\ref{leq.1}) by neglecting terms that would change ${\cal N}$.  Accordingly, we resort to the 
 approximate equation for $\rho(t)$ given by 
 
 \begin{widetext}
  \begin{eqnarray}
\frac{\partial \rho (t)}{ \partial t}  &\approx&   - i \{ H_{\rm MF} (t) \rho(t) - \rho (t) H_{\rm MF} (t) \} 
 -  g \sum_{\bf k} \{ [\Gamma_{{\bf k},+} (t)]^\dagger \Gamma_{{\bf k},+}(t) \rho (t) \nonumber \\
  &+&  
 \rho (t)  [\Gamma_{{\bf k},+} (t)]^\dagger \Gamma_{{\bf k},+}(t)
 + \Gamma_{{\bf k},-} (t) [\Gamma_{{\bf k},-} (t)]^\dagger \rho (t) + 
 \rho (t )  \Gamma_{{\bf k},-} (t) [\Gamma_{{\bf k},-} (t)]^\dagger \}
 \;\;\;\; . 
 \label{smalln.4}
 \end{eqnarray}
 \noindent 
 \end{widetext}
 As a result, retaining only the matrix elements of $\rho(t)$ between states at half-filling
 (that is, states containing ${\cal N}=N$ particles, due to the spin degeneracy), 
 we write it in the approximate form

 \begin{eqnarray}
 \rho (t) &\approx& \sum_{\{ {\bf q}_j \} } \sum_{\{ \lambda_j \} ; \{ \mu_j \} } 
  \rho^{(  N )}_{ \{\lambda_j \} , \{ \mu_j \}  , \{ {\bf q}_j \}} (t) \times   \label{smalln.3x} \\
 &&   | N,  t \rangle_{\{{\bf q}_j \},\{\lambda_j\}} \; ~_{\{{\bf q}_j \} , \{ \mu_j \}}\langle N,t | \;
 \:, \nonumber 
 \end{eqnarray}
 \noindent 
 with $N$ being the number of lattice sites. 
 
 Next, we note that, due to the parametric dependence on $t$ of the operators $\Gamma_{{\bf k},\lambda} (t)$, 
a solution of the time-dependent Schr\"odinger equation 

\beq
\frac{ \partial }{\partial t} | \psi (t ) \rangle = H_{\rm MF} (t ) | \psi (t ) \rangle 
\; , 
\label{smalln.y1}
\eneq
\noindent
is not simply provided by setting 

\beq
|\psi (t) \rangle =  \exp \left[-i\int_0^t \: \sum_{j=1}^{N} \lambda_j \epsilon_{{\bf q}_j } ( \tau) 
 \: d \tau \right] \:    | N,  t \rangle_{\{{\bf q}_j \},\{\lambda_j\}}
 \;  , 
 \label{smallin.6}
 \eneq
 \noindent
 as one would in fact obtain

 \begin{eqnarray}
&& \frac{ \partial }{\partial t}    |\psi (t)\rangle  = 
-i H_{\rm MF}(t)  | \psi (t)  \rangle   + 
 \label{smallin.7}\\
&&  \exp \left[-i\int_0^t \: \sum_{j=1}^{N} \lambda_j \epsilon_{{\bf q}_j } ( \tau) 
 \: d \tau \right] \:  \frac{\partial}{\partial t}    | N,  t \rangle_{\{{\bf q}_j \},\{\lambda_j\}}
 \; . \nonumber 
 \end{eqnarray}
 \noindent
 Yet,  while the ``dynamical'' phases at the right-hand side of Eq.(\ref{smallin.6})
 typically grow linearly with time $t$, the time evolution of the state 
 $|N,t \rangle_{\{{\bf q}_j \},\{\lambda_j\}}$ (which is determined by the 
 parametric dependence on $t$ of the operators $\Gamma_{{\bf k},\lambda} (t)$), 
 takes place over periodic patterns in time. For this reason, 
   it is reasonable to assume that the dependence on time of the dynamical phases   
 takes place over typical frequencies much larger than the 
 one associated to the parametric dependence  of $|N,t \rangle_{\{{\bf q}_j \},\{\lambda_j\}}$ on $t$. 
 Thus, in the following we neglect the latter contribution to the right-hand side of 
 Eq.(\ref{smallin.7}). This leads us to write  a simplified (and closed) set of equations for the 
 matrix elements $  \rho^{(N)}_{\{\lambda_j \},\{ \mu_j \},\{ {\bf q}_j \}} (t) $, given by
  
 \begin{eqnarray}
&& \frac{  \rho^{(N)}_{\{\lambda_j \},\{ \mu_j \},\{ {\bf q}_j \}} (t) }{ d t } = \nonumber \\
&& - i \sum_{j=1}^N \{ [ \lambda_j - \mu_j ] \epsilon_{{\bf q}_j} (t) \}  \rho^{(N)}_{\{\lambda_j \},\{ \mu_j \},\{ {\bf q}_j \}} (t) 
 \nonumber \\
&&  - g \{2 N + \sum_{j=1}^N [ \lambda_j + \mu_j ]\}  \rho^{(N)}_{\{\lambda_j \},\{ \mu_j \},\{ {\bf q}_j \}} (t) 
   \:  . 
 \label{smallin.9}
 \end{eqnarray}
 \noindent
 Upon integrating Eqs.(\ref{smallin.9}), we obtain 
 
 \begin{eqnarray}
&& \rho^{(N)}_{\{\lambda_j \},\{ \mu_j \},\{ {\bf q}_j \}} (t) = e^{ \left[ - i \int_0^t \: d \tau \: 
  \sum_{j=1}^N [\lambda_j - \mu_j ] \epsilon_{{\bf q}_j} ( \tau ) \right] } \times
 \label{smallin.10}\\
&& \exp \left[  -  g \{  2 N + \sum_{j=1}^N [ \lambda_j + \mu_j ] \}  t \right]   
 \rho^{(N)}_{\{\lambda_j \},\{ \mu_j \},\{ {\bf q}_j \}} (0)  \nonumber 
  \:  . 
 \label{smallin.10}
 \end{eqnarray}
 \noindent
 An important consequence of Eq.(\ref{smallin.10}) is that all the   elements 
$\rho^{N}_{  \{\lambda_j \} , \{ \mu_j \} , \{ {\bf q}_j \}  } (t) $ are exponentially suppressed, as soon as
$2gt \geq 1$, except for the diagonal ones with $\lambda_1 = \ldots = \lambda_N = -$, and 
$\mu_1 = \ldots = \mu_N = -$. Over time scales $> (2g)^{-1}$, we therefore obtain 

\beq
 \rho (t ) \approx   | N,t \rangle_{\{{\bf q}_j\},\{-\}} \;  ~_{\{{\bf q}_j\},\{-\}} \langle   N,t | \: 
 \:.
 \label{smallin.13}
 \eneq
 \noindent
 Moreover,  
 we point out how,  in writing the right-hand side of Eq.(\ref{smallin.13}), we did not sum over the ${\bf q}_j$, as 
 the state  $ | N,t \rangle_{\{{\bf q}_j\},\{-\}}$ is uniquely fixed   by populating the negative-energy modes at time $t$
  for all possible values of ${\bf q}_j$. As a result of our approximations, we eventually find

 \begin{eqnarray}
&& {\cal F} (t ) = \langle \psi (0) | \rho ( t ) | \psi (0) \rangle\nonumber \\
&& \approx  
 | \langle  \psi (0) | N,t \rangle_{\{{\bf q}_j\},\{-\}} |^2 
 \:  . 
 \label{smallin.14}
 \end{eqnarray}
 \noindent
 Remarkably,  Eq.(\ref{smallin.14}), which is valid for $2gt > 1$ and which 
 provides us with  the starting point of  our following derivation, 
   coincides with the 
 value that the Loschmidt echo would have in a closed system whose (pure) collective
 state, at time $t>0$,  is  given by $ | \psi (t) \rangle = \prod_j [ \Gamma_{{\bf q}_j , -}(t)]^\dagger  |{\bf 0} \rangle$. 
 In fact, the analogy is not accidental.  For a closed system, the  Loschmidt echo
 is nothing but a  fidelity between the state at the initial time $t= 0$ and its
time-evolved counterpart at general $t$. Therefore, if the evolved state crosses a quantum phase transition, 
a nonanalyticity is expected on ${\cal F}(t)$ \cite{Zanardi2006_2,Zanardi2006,Heyl2019}. 

\begin{figure}
 \center
\includegraphics*[width=.9 \linewidth]{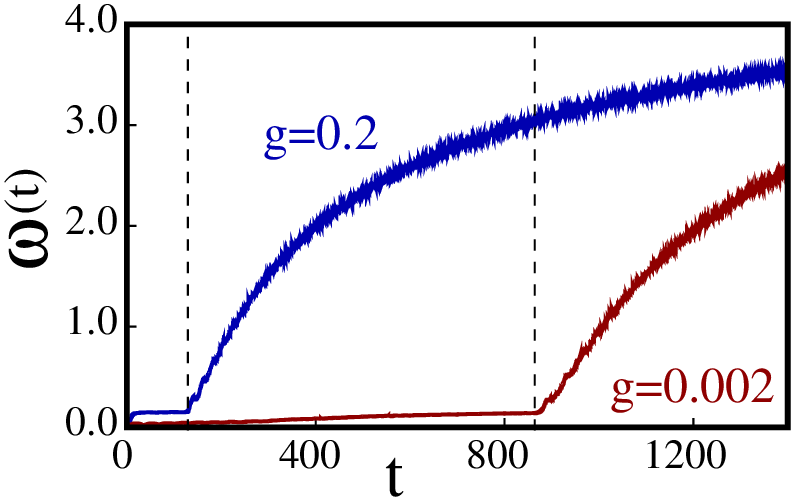}
\caption{ Rate function  $\omega (t)$ (Eq.(\ref{omem.1}))  as a function of time $t$ computed with the time-dependent 
MF Hamiltonian with parameters $\Delta_S (t)$ and $\Delta_{xy} (t)$ as in 
Fig.\ref{sid_f}, using $g=0.2$ (blue curve) and $g=0.002$ (red curve).   
The dashed  vertical lines mark the DPT in the two cases.}
\label{echo_sid}
\end{figure}

To probe the DPT, in the following we rather look  for nonanalyticities
  in the  rate function $\omega (t)$,  defined as \cite{Jurcevic2017,Zvyagin2016,Heyl2018,Heyl2019,shorter_paper} 
   
  \beq
  \omega (t) = - \frac{1}{N} \log [ {\cal F} (t)]
  \;\;,
  \label{omem.1}
  \eneq
  \noindent 
 by computing ${\cal F} (t)$ as 

\beq
{\cal F} (t) = | \langle \psi (t=0 ) | {\cal U} (t) | \psi (t=0)\rangle |^2
\;\;\;\; , 
\label{lec.1}
\eneq
\noindent
with ${\cal U} (t ) = {\cal T} \exp \left[ -i \int_0^t d \tau H ( \tau ) \right]$, where ${\cal T}$ is the time-ordered
evolution operator.  To compute the right-hand side of Eq.(\ref{lec.1}),  we follow a two-step procedure. Specifically, 
we   first numerically compute $\Delta_{\bf k} ( t )$ within the time-dependent SCMF approximation. 
Therefore, we use $\Delta_{\bf k} (t)$ 
self-consistently computed as an input parameter of the time-dependent Hamiltonian  $H_{\rm MF} (t)$, which we eventually employ  
 to compute the right-hand side of Eq.(\ref{lec.1}). In this way, we   compute $\omega (t)$ along the time evolution of 
 the systems  with  parameters set as in drawing Fig.\ref{sid_f}.  In Fig.\ref{echo_sid} we draw the corresponding  plot of
 $\omega (t)$. The blue and the red curve respectively 
 correspond  to $g=0.2$ and to $g=0.002$, with all the other parameters chosen exactly as  in 
Fig.\ref{sid_f}.  In both cases we mark with a vertical  dashed line the time $t=t_*$ at which
 the system goes through the DPT. Despite some differences between the two plots, including, of 
 course, the different values for $t_*$ determined by the different values of $g$,   we 
 note an over-all  similar behavior of $\omega (t)$.  
Specifically, for $0 \leq t < t_*$, $\omega (t)$ takes only a mild time dependence  on $t$, with  
 $\omega (t) \sim 0.1-0.2$, denoting an 
appreciable overlap between $|\psi (0)\rangle$ and $|\psi (t)\rangle$. Therefore, we see that 
the first part  of the plots indicate the persistence of the system in the initial pre-quench phase for
 times $t$ up to the transition time $t_*$  \cite{Heyl2019}. At $t=t_*$,  a sudden change in the slope of 
 $\omega (t)$  evidences how $t=t_*$ corresponds to a point where the derivative of 
$\omega (t)$ does not exist, that is, to a typical sort of nonanalyticity that signals a 
DPT. For $t>t_*$, the rapid increase in $\omega (t)$,  following the sudden change in the slope, corresponds
to a drastic reduction in ${\cal F} (t)$ (by orders of magnitude), which is a clear signal that, moving across $t=t_*$, the system 
has gone through a DPT.

\noindent
 About the relation between $t_*$ and the coupling $g$ we note that 
the  physical  intuition behind  the existence of a critical time for a DPT  is related to 
the  geometric properties of the energy landscape of the system \cite{Lu2017}. During the dissipative 
dynamics induced by the coupling with the bath, the system evolves with a speed that depends on its geometric properties. 
If the system crosses a ``flat''   region in energy, the time evolution is extremely slow. 
As soon as the edge of the stationary solution is reached, the evolution becomes extremely fast and the DPT 
toward the true stationary solution takes place. 
The critical time at which this happens depends on the trajectory itself and can not easily be predicted 
due to the fact that the energy landscape itself 
is a function of time in the SCMF. A similar behavior has been observed in a much simpler spin system, 
where it has also been observed that $t_*$  can depend on the existence of shortcuts in the energy 
landscapes \cite{Nava2019}, or on the values of the 
bath-system coupling strengths \cite{Nava2022}.  

 While we do not discuss this point in our paper, it is finally worth mentioning that, in addition to the 
fidelity, one might also potentially use the entropy $S(t)$ as an effective mean to detect the DPT. 
Indeed, along the derivation presented in Ref. \cite{Bacsi2023}, we expect that, in the zero-temperature limit, 
$S(t)$ for our system would be 0 both at $t=0$ (because our system is prepared in a pure state), as well 
as for $t\to \infty$ (because asymptotically our system is described by a Boltzmann distribution at $T=0$). 
In between, for $g=0.2$, from the plot of  Fig.\ref{sid_f}{\bf a)}, we infer that the time evolution of the system is characterized 
by large intervals of time over which the gaps keep constant, and by rapid changes in the gap themselves right after
starting the time evolution and at the DPT. The rapid changes in the gaps can be effectively regarded as 
quenches of the superconducting order parameters. Therefore, 
by analogy with what is discussed in Ref.\cite{Bacsi2023} for a bosonic system, we expect that a significant number 
of quasiparticle excitations are created at any change in the gaps, contributing to a corresponding sharp increase of the entropy.
The increase of the entropy should, therefore, work as a signal of the DPT. At smaller values of $g$, we expect that the
entropy increase is present at the DPT, as well, although the feature should be  smoother and less marked.
  
\section{Discussion and Conclusions}
\label{conclusions}

In this paper we have constructed a protocol to induce   nonequilibrium dynamics in an open, superconducting 
system coupled to an 
external bath. Pertinently choosing the jump operators in the  Lindblad master equation  approach to the dissipative 
dynamics of the density matrix of the system,  we  let the  system evolve toward 
the thermodynamical stationary state, by making sure that  
 the Boltzmann distribution is a stationary solution of the Lindblad equation.
 Along our derivation, we have discussed in 
detail how the mismatch between the initial state and the asymptotic state of the system can lead to a dynamical 
phase transition, which, under suitable conditions, may also determine a transition between a topologically
nontrivial and a topologically trivial phase, or vice versa \cite{shorter_paper}.
 
To monitor the system across the  DPT, we look at  the self-consistently computed 
 superconducting gap $\Delta_{\bf k} (t)$ and  at the fidelity ${\cal F} (t)$. At the time $t_*$ at which the phase 
 transition takes place,  the components of $\Delta_{\bf k} (t)$ abruptly change: this corresponds 
 to a nonanalyticity (a change in the slope) of the function $\omega (t)=-\frac{\ln {\cal F} (t)}{N}$, that is, 
 a point where $\omega (t)$ is not differentiable.  
 
 As a general comment we note that,  while there is already a remarkable amount of results on DPTs in closed systems,  
 still very little is known about DPTs in open systems.    In our paper, we attempt to fill such a gap by performing an explicit model 
 calculation of a DPT in  superconducting, open systems. Among the results we obtain along our derivation we show how, in an 
 open system,  the mismatch between the initial state and  the choice of the Hamiltonian parameters, combined with the relaxation 
 dynamics due to the coupling to the bath, triggers the onset of the DPT,  how the location in time of the DPT ($t_*$) is 
 affected by the coupling to the bath, and how it is possible, by pertinently tuning the system  parameters, to select the asymptotic state toward which the system evolves.  
  
In principle, our approach can be readily generalized to a generic dynamical phase transition in other 
many-body, fermionic systems \cite{Guerci2021,Giuliano2020,Giuliano2020a}.   Of course, our model is amenable to
substantial improvements, possibly
on the numerical computational side, such as resorting to a fully time-dependent mean field 
Hamiltonian $H_{\rm MF} (t)$, in which $\Delta_{\bf k} (t)$, self-consistently computed, should appear as 
a time-dependent parameter. Also,  it would be extremely interesting to perform a systematic analysis of 
 how the critical time $t_*$ depends on the value of $g$, thus  to eventually recover the results 
of Ref. \cite{Peronaci2015} as a limiting case of ours. While interesting, all these tasks fall beyond the scope of this paper, and 
we are planning to address them as a further development  of 
the work we present here.

 \vspace{0.3cm} 
 
{\bf Acknowledgements:}   We thank N. Lo Gullo and F. Plastina for insightful discussions. A.N., C.A.P., L.L., and D.G.  
  acknowledge   financial support  from Italy's MIUR  PRIN project  TOP-SPIN  (Grant No. PRIN 20177SL7HC).   
  L.L.  acknowledges financial support by a project funded under the National Recovery and Resilience Plan (NRRP), Mission 4 Component 2 Investment 1.3 - Call for tender No. 341 of 15/03/2022 of Italian Ministry of University and Research funded by the European Union – NextGenerationEU, award number PE0000023, Concession Decree No. 1564 of 11/10/2022 adopted by the Italian Ministry of University and Research, CUP D93C22000940001, Project title "National Quantum Science and Technology Institute" (NQSTI). 
 A.N. and R.E. acknowledge  funding by the Deutsche Forschungsgemeinschaft (DFG, German Research Foundation) under Grant 
 No.~ 277101999, TRR 183 (project C01), under Germany's Excellence Strategy - Cluster of Excellence Matter
 and Light for Quantum Computing (ML4Q) EXC 2004/1 - 390534769, and under Grant No.~EG 96/13-1.

\appendix 

\section{Self-consistent mean-field approximation for the superconducting Hamiltonian in Eq.(\ref{eh.1a})}
\label{mhe}

In this appendix we provide the details of the SCMF approximation, through which we trade  $H$ in Eq.(\ref{eh.1a}) for 
the MF Hamiltonian, $H_{\rm MF}$  in Eq.(\ref{eh.2a}). 

In the  Hamiltonian of Eq.(\ref{eh.1a}) we have introduced three different 
interactions, which, in resorting to the SCMF approximation, we decouple
as follows:

\begin{itemize}

\item {\it Local superconducting pairing:}

\begin{eqnarray}
&& - U\sum_{\bf r} \langle c_{{\bf r},\uparrow}^\dagger c_{{\bf r},\uparrow} c_{{\bf r},\downarrow}^\dagger c_{{\bf r},\downarrow} \rangle  
 \to  U \sum_{\bf r}\langle c_{{\bf r},\downarrow} c_{{\bf r},\uparrow} \rangle \langle c_{{\bf r},\uparrow}^\dagger c_{{\bf r},\downarrow}^\dagger \rangle
\nonumber \\
&&-U\sum_{\bf r} \langle c_{{\bf r},\downarrow} c_{{\bf r},\uparrow} \rangle c_{{\bf r},\uparrow}^\dagger c_{{\bf r},\downarrow}^\dagger  
- U \sum_{\bf r} c_{{\bf r},\downarrow} c_{{\bf r},\uparrow} \langle c_{{\bf r},\uparrow}^\dagger c_{{\bf r},\downarrow}^\dagger   \rangle
= \nonumber \\
&& \frac{N}{U} | \Delta_S  |^2 - \sum_{{\bf r}} \{ \Delta_S c_{{\bf r},\uparrow}^\dagger c_{{\bf r},\downarrow}^\dagger  
+  \Delta_S^*    c_{{\bf r},\downarrow} c_{{\bf r},\uparrow} \}
\:\:\:, 
\label{eh.2c}
\end{eqnarray}
\noindent
with  
$\Delta_S = U  \langle c_{{\bf r},\downarrow} c_{{\bf r},\uparrow} \rangle$.

\item {\it Nearest-neighbor superconducting pairing:}

\begin{eqnarray}
&&  - \frac{V}{2}\sum_{{\bf r},\hat{\delta}} \sum_\sigma \sum_{\sigma'} c_{{\bf r},\sigma}^\dagger  c_{{\bf r},\sigma}  
c_{{\bf r} + \hat{\delta},\sigma'}^\dagger  c_{{\bf r}+ \hat{\delta},\sigma'}  \nonumber \\
&& \to \frac{V}{2} \sum_{\bf r}\sum_{\hat{\delta}}\sum_\sigma 
\langle c_{{\bf r},\sigma} c_{{\bf r}+\hat{\delta},\bar{\sigma}}\rangle \langle c_{{\bf r}+\hat{\delta},\bar{\sigma}}^\dagger 
c_{{\bf r},\sigma}^\dagger  \rangle
\nonumber \\
&& -\frac{V}{2}  \sum_{\bf r}\sum_{\hat{\delta}}\sum_\sigma 
\langle c_{{\bf r},\sigma} c_{{\bf r}+\hat{\delta},\bar{\sigma}}\rangle c_{{\bf r}+\hat{\delta},\bar{\sigma}}^\dagger 
c_{{\bf r},\sigma}^\dagger \nonumber \\
&& - \frac{V}{2} \sum_{\bf r}\sum_{\hat{\delta}}\sum_\sigma 
c_{{\bf r},\sigma} c_{{\bf r}+\hat{\delta},\bar{\sigma}}  \langle c_{{\bf r}+\hat{\delta},\bar{\sigma}}^\dagger 
c_{{\bf r},\sigma}^\dagger  \rangle = \nonumber \\
&& \frac{N}{V} \sum_{\hat{\delta}}  |\Delta_{NN} (\hat{\delta} )|^2  
-  \sum_{\bf r}\sum_{\hat{\delta}}  \{ \Delta_{NN} (\hat{\delta} )   c_{{\bf r}+\hat{\delta},\uparrow}^\dagger 
c_{{\bf r},\downarrow}^\dagger \nonumber \\
&&  +[ \Delta_{NN} (\hat{\delta} )]^*  
c_{{\bf r},\downarrow} c_{{\bf r}+\hat{\delta},\uparrow}  \} 
 \:\:\:,
\label{eh.3c}
\end{eqnarray}
\noindent
with $\Delta_{NN} (\hat{\delta})=V\langle c_{{\bf r},\downarrow} c_{{\bf r}+\hat{\delta},\uparrow} \rangle$ and 
with the additional assumption that $\langle c_{{\bf r},\downarrow} c_{{\bf r}+\hat{\delta},\uparrow} \rangle= 
- \langle c_{{\bf r},\uparrow} c_{{\bf r}-\hat{\delta},\downarrow} \rangle$. (Here, $\bar\sigma=-\sigma$ 
is the spin index opposite to $\sigma$.)

\item {\it Next-to-nearest-neighbor superconducting pairing:}

\begin{eqnarray}
&&  - \frac{Z}{2}\sum_{{\bf r},\hat{\delta}'} \sum_\sigma \sum_{\sigma'} c_{{\bf r},\sigma}^\dagger  c_{{\bf r},\sigma}  
c_{{\bf r} + \hat{\delta}',\sigma'}^\dagger  c_{{\bf r}+ \hat{\delta}',\sigma'} \nonumber \\
&&  \to \frac{Z}{2} \sum_{\bf r}\sum_{\hat{\delta}'}\sum_\sigma 
\langle c_{{\bf r},\sigma} c_{{\bf r}+\hat{\delta}',\bar{\sigma}}\rangle \langle c_{{\bf r}+\hat{\delta}',\bar{\sigma}}^\dagger 
c_{{\bf r},\sigma}^\dagger  \rangle
\nonumber \\
&& -\frac{Z}{2}  \sum_{\bf r}\sum_{\hat{\delta}}\sum_\sigma 
\langle c_{{\bf r},\sigma} c_{{\bf r}+\hat{\delta}',\bar{\sigma}}\rangle c_{{\bf r}+\hat{\delta}',\bar{\sigma}}^\dagger 
c_{{\bf r},\sigma}^\dagger \nonumber \\
&& - \frac{Z}{2} \sum_{\bf r}\sum_{\hat{\delta}}\sum_\sigma 
c_{{\bf r},\sigma} c_{{\bf r}+\hat{\delta}',\bar{\sigma}}  \langle c_{{\bf r}+\hat{\delta}',\bar{\sigma}}^\dagger 
c_{{\bf r},\sigma}^\dagger  \rangle = \nonumber \\
&& \frac{N}{Z} \sum_{\hat{\delta}}  |\Delta_{NNN} (\hat{\delta}' )|^2  
-  \sum_{\bf r}\sum_{\hat{\delta}}  \{ \Delta_{NNN} (\hat{\delta}' )   c_{{\bf r}+\hat{\delta}',\uparrow}^\dagger 
c_{{\bf r},\downarrow}^\dagger  \nonumber \\
&& +[ \Delta_{NNN} (\hat{\delta}' )]^*  
c_{{\bf r},\downarrow} c_{{\bf r}+\hat{\delta}',\uparrow}  \} 
 \:\:\:,
\label{eh.3b}
\end{eqnarray}
\noindent
with $\Delta_{NNN} (\hat{\delta}')=Z\langle c_{{\bf r},\downarrow} c_{{\bf r}+\hat{\delta}',\uparrow} \rangle$ and 
with the additional assumption that  $\langle c_{{\bf r},\downarrow} c_{{\bf r}+\hat{\delta}',\uparrow} \rangle= 
- \langle c_{{\bf r},\uparrow} c_{{\bf r}-\hat{\delta}',\downarrow} \rangle$.
\end{itemize}
Resorting to Fourier space, we   obtain  
$H=H_K+H_P + H_Q$, with the kinetic energy and  the pairing term respectively given by 

\begin{widetext}
\begin{eqnarray}
H_K&=&\sum_{{\bf k}} \sum_\sigma \{ -2 [\cos (k_x )+ \cos (k_y ) ] - 2t' [\cos (k_x+k_y)+\cos (k_x-k_y)] - \mu \} c_{{\bf k},\sigma}^\dagger 
c_{{\bf k},\sigma} \equiv \sum_{{\bf k}} \sum_\sigma \xi_{{\bf k}} c_{{\bf k},\sigma}^\dagger c_{{\bf k},\sigma} \nonumber \\
H_P &=&- \sum_{{\bf k}} \{\Delta_S + \sum_{\hat{\delta}} e^{-i{\bf k}\cdot \hat{\delta}} \Delta_{NN}(\hat{\delta} ) 
+  \sum_{\hat{\delta}'} e^{-i{\bf k}\cdot \hat{\delta}'} \Delta_{NNN}(\hat{\delta}' ) \} c_{{\bf k},\uparrow}^\dagger c_{{\bf - k},\downarrow}^\dagger 
+{\rm h.c.}
\nonumber \\
 &\equiv& - \sum_{{\bf k}} \{ \Delta_{\bf k} c_{{\bf k},\uparrow}^\dagger c_{{\bf -k},\downarrow}^\dagger + 
 [ \Delta_{\bf k}]^* c_{{\bf -k},\downarrow}  c_{{\bf k},\uparrow} \}
    \:\: ,
\label{eh.5c}
\end{eqnarray}
\noindent
\end{widetext}
 with
 
 \begin{eqnarray}
 \xi_{\bf k}&=& -2 [\cos (k_x )+ \cos (k_y ) ] - 4t' \cos(k_x)\cos(k_y) - \mu \nonumber \\
 \Delta_{\bf k}&=& \Delta_S+ \sum_{\hat{\delta}} e^{-i{\bf k}\cdot \hat{\delta}} \Delta_{NN}(\hat{\delta} ) 
+  \sum_{\hat{\delta}'} e^{-i{\bf k}\cdot \hat{\delta}'} \Delta_{NNN}(\hat{\delta}' )
\: . \nonumber 
\end{eqnarray}
\noindent
Setting 

\begin{eqnarray}
\Delta_{NN} (\hat{\delta})&=&\Biggl\{ \begin{array}{l} +\Delta_{x^2-y^2} \;\; , \; {\rm if} \: \hat{\delta}=\pm \hat{x} \\
-\Delta_{x^2-y^2} \;\; , \; {\rm if} \: \hat{\delta} = \pm \hat{y} 
\end{array} \nonumber \\
\Delta_{NNN} (\hat{\delta}') &=&\Biggl\{ \begin{array}{l} -i\Delta_{xy} \;\; , \; {\rm if} \; \hat{\delta}' = \pm (\hat{x}+\hat{y}) \\
+ i\Delta_{xy} \;\; , \; {\rm if} \; \hat{\delta}'=\pm (\hat{x}-\hat{y}) \end{array}
\;\; , 
\label{eh.9b}
\end{eqnarray}
\noindent
we obtain the expression of $\Delta_{\bf k}$ in Eq.(\ref{eh.4a}). 
Finally,  the  energy of the superconducting condensate, $H_Q$, is 
given by  

\beq
H_Q=\frac{N}{U}|\Delta_S|^2 + \frac{4 N}{V}  |\Delta_{x^2-y^2}  |^2 
+ \frac{4 N}{Z}  |\Delta_{xy} |^2 
\:\:\: . 
\label{eh.7}
\eneq
\noindent
Minimizing the total energy with respect to $\Delta_S,\Delta_{x^2-y^2}$, and $\Delta_{xy}$, 
we obtain  the self-consistent
equations for the gap order parameter, given by   

\begin{eqnarray}
\Delta_S &=& \frac{U}{2 N} \sum_{\bf k}  \frac{\Delta_S}{\epsilon_{\bf k}}  \varphi  ( \epsilon_{\bf k}) 
  \nonumber \\
\Delta_{x^2-y^2} &=& \frac{V}{ 2  N} \sum_{\bf k} \frac{\Delta_{x^2-y^2} [\cos(k_x)-\cos(k_y)]^2 }{\epsilon_{\bf k}}  
 \varphi  ( \epsilon_{\bf k}) \nonumber \\
\Delta_{xy} &=& \frac{2Z}{  N} \sum_{\bf k}  \frac{\Delta_{xy} \sin^2(k_x)\sin^2(k_y) }{\epsilon_{\bf k}}  
 \varphi  ( \epsilon_{\bf k}) 
\:\:\: ,
\label{sc.13}
\end{eqnarray}
\noindent
with $\epsilon_{\bf k}=\sqrt{\xi_{\bf k}^2 + |\Delta_{\bf k}|^2}$ and 
$\varphi ( \epsilon_{\bf k}) =  f(-\epsilon_{\bf k})-f(\epsilon_{\bf k})$, 
with $f(\epsilon_{\bf k})$ being Fermi distribution function. 

From the self-consistent equations in Eqs.(\ref{sc.13}) (taken in the zero-temperature limit) we have derived the phase diagram 
discussed in the main text.

\section{Relaxation dynamics following a sudden quench on $\Delta_{\bf k} (t)$}
\label{sudden}

In this appendix we  present a simplified version of the approach we used throughout our paper. Specifically, 
rather than quenching, at $t=0$, the interaction strengths, we directly quench the superconducting order 
parameter, so that it takes the form

\beq
\hat{\Delta}_{\bf k} (t) = \Delta^{(0)}_{\bf k} \theta (-t) + \Delta_{\bf k} \theta (t)
\:\:\:\: . 
\label{sudden.1}
\eneq
\noindent
As a result of giving up self consistency,  for $t>0$   Eqs.(\ref{leq.3}) become purely linear and 
simplify to
 
\begin{eqnarray}
\frac{d \nu_{\bf k} (t)}{dt}&=& - 2 g \nu_{\bf k} ( t) +  \frac{g \xi_{\bf k}}{\epsilon_{\bf k}}  + 2 \Im m [  \Delta_{\bf k} [f_{\bf k} (t) ]^* ]
\nonumber \\ 
\frac{ d f_{\bf k} (t)}{dt} &=& -2 ( g + i \xi_{\bf k} )f_{\bf k}(t) +2  i  \Delta_{\bf k} \nu_{\bf k} (t) +    \frac{g \Delta_{\bf k}}{\epsilon_{\bf k}} 
\: , 
\label{sudden.2}
\end{eqnarray}
\noindent
with   $\epsilon_{\bf k} = \sqrt{\xi_{\bf k}^2 + | \Delta_{\bf k}|^2}$ and 
the initial conditions given by 

\begin{eqnarray}
\nu_{\bf k} (t=0) &=& \frac{\xi_{\bf k}}{\epsilon_{\bf k}^{(0)} } \nonumber \\
f_{\bf k} (t=0 ) &=& \frac{\Delta_{\bf k}^{(0)}}{\epsilon_{\bf k}^{(0)} } 
\:\:\:\: , 
\label{sudden.x1}
\end{eqnarray}
\noindent
with $\epsilon_{\bf k}^{(0)} = \sqrt{\xi_{\bf k}^2 + | \Delta^{(0)}_{\bf k}|^2}$. We may now readily solve 
Eqs.(\ref{sudden.2}) in terms of the Laplace transforms of $\nu_{\bf k}(t)$ and $f_{\bf k}(t)$. 
As a result, we obtain 
 
 \begin{widetext}

\begin{eqnarray}
\nu_{\bf k} (z) &=& \frac{g\xi_{\bf k}}{\epsilon_{\bf k} z (z+2g)} + 
\left[ \frac{(z+2g) \{ 2 \Im m \{ [ f_{\bf k} (0) ]^* \Delta_{\bf k} \}  + (z+2g) \nu_{\bf k}(0) \} 
+ 4  \xi_{\bf k} \Re e\{ [f_{\bf k} (0)]^* \Delta_{\bf k} \} + 4 \nu_{\bf k} ( 0 ) \xi_{\bf k}^2 }{ (z+2g) 
[ (z+2g)^2 + 4 \epsilon_{\bf k}^2] } \right] 
\label{sudden.x3} \\
f_{\bf k} (z) &=&  \frac{g\Delta_{\bf k}}{\epsilon_{\bf k} z (z+2g)} +
 \left[ \frac{f_{\bf k}(0) (2 | \Delta_{\bf k}|^2 + (z+2g)(z+2g -2i\xi_{\bf k}) )
 + 2 \Delta_{\bf k} ( \Delta_{\bf k} [f_{\bf k} (0)]^*  + i \nu_{\bf k}(0) (z+2g-2i\xi_{\bf k} ) ) }{ (z+2g) 
[ (z+2g)^2 + 4 \epsilon_{\bf k}^2] } \right]
\:. \nonumber 
\end{eqnarray}
\noindent
\end{widetext}
In the three cases  we are investigating here, the 
Laplace transforms of the superconducting gap,  
$\Delta_S (z) , \Delta_{x^2-y^2} (z)$, and $\Delta_{xy} (z)$, are given by 

\begin{eqnarray}
\Delta_S (z) &=& \frac{U}{2N} \sum_{\bf k} f_{\bf k} (z) \nonumber \\
\Delta_{x^2-y^2} (z) &=& \frac{V}{2N} \sum_{\bf k} \{ \cos (k_x ) - \cos (k_y ) \} f_{\bf k} (z) \nonumber \\
\Delta_{xy} (z) &=& \frac{2iZ}{2N} \sum_{\bf k} \sin (k_x ) \sin (k_y ) f_{\bf k} (z) 
\:\:\:\: . 
\label{sudden.x4}
\end{eqnarray}
\noindent
Using Eqs.(\ref{sudden.x3}) and going through Eqs.(\ref{sudden.x4}), we can readily compute the 
position of the poles of the Laplace transforms of the superconducting gaps, which provide us with 
the relevant informations concerning the gap dynamics. To do so, we first of all 
replace $\nu_{\bf k}(0)$ and  $f_{\bf k}(0)$  with their expressions in Eqs.(\ref{sudden.x1}), 
by setting 

\begin{eqnarray}
\Delta_{\bf k}^{(0)} &=& \Delta_S^{(0)} + 2 \Delta_{x^2-y^2}^{(0)}  [ \cos (k_x) - \cos (k_y ) ] \nonumber \\
&-& 4i\Delta_{xy}^{(0)} \sin (k_x ) \sin (k_y ) 
\nonumber \\
\epsilon_{\bf k}^{(0)} &=& \sqrt{ \xi_{\bf k}^2 + | \Delta_{\bf k}^{(0)}|^2 }
\:\:\:\: . 
\label{sudden.y1}
\end{eqnarray}
\noindent
 Moreover, we also set 
 
 \begin{eqnarray}
 \Delta_{\bf k} (z) &=& \Delta_S (z)+ 2 \Delta_{x^2-y^2} (z)[ \cos (k_x) - \cos (k_y ) ] \nonumber \\
 &-& 4i\Delta_{xy} (z) \sin (k_x ) \sin (k_y ) 
 \:\:\:\: . 
 \label{sudden.y2}
 \end{eqnarray}
  \noindent
 From the explicit expression of $f_{\bf k} (z)$ in Eqs.(\ref{sudden.x3}) we can infer the relaxation dynamics of 
 the superconducting order parameter for $t \geq0$. Indeed, 
 we readily identify two single poles at $z=0$ and at $z=-2g$. The former one determines  the 
 asymptotic value of the superconducting gap. Taking the corresponding residue and employing the time-dependent 
 version of Eqs.(\ref{sudden.x4}), we readily find that, from the pole at $z=0$, the superconducting order 
 parameter as $t \to \infty$ takes a contribution equal to the after-the-quench value. An additional simple pole 
 takes place at $z=-2g$, which corresponds to a damping of the corresponding contribution to 
 $\Delta_{\bf k} (t)$ as $e^{-2gt}$. 
 Finally, an additional complex pole is expected to arise at $z=-2g+i\omega_*$, with $\omega_*$ determined by the integration
 over $d^2 k$: this determines again  an exponential damping of the corresponding contribution to the superconducting gap 
 over a time scale $ \sim (2g)^{-1}$ on top of an oscillating modulation with frequency $\omega_*$. Apparently, as long as 
 Finally, an additional complex pole is expected to arise at $z=-2g+i\omega_*$, with $\omega_*$ determined by the integration
 over $d^2 k$: this determines again  an exponential damping of the corresponding contribution to the superconducting gap 
 over a time scale $ \sim (2g)^{-1}$ on top of an oscillating modulation with frequency $\omega_*$. Apparently, as long as 
  $g>0$, all the contributions are washed out by the exponential damping, except 
  the ones entering the after-the-quench $\Delta_{\bf k}$, according to Eq. (\ref{sudden.y2}). As $g \to 0$ the  asymptotic
  behavior becomes   more involuted, also depending on the symmetry of the order 
  parameter.  From the above discussion, we expect that, when only a single interaction strength is 
  different from zero, the relaxation time scale of the corresponding order parameter
  is independent of its symmetry as, in fact, witnessed by the results  in  Fig. \ref{quench_single}.

\bibliography{lindblad_super}

%merlin.mbs apsrev4-1.bst 2010-07-25 4.21a (PWD, AO, DPC) hacked
%Control: key (0)
%Control: author (8) initials jnrlst
%Control: editor formatted (1) identically to author
%Control: production of article title (-1) disabled
%Control: page (0) single
%Control: year (1) truncated
%Control: production of eprint (0) enabled
\begin{thebibliography}{65}%
\makeatletter
\providecommand \@ifxundefined [1]{%
 \@ifx{#1\undefined}
}%
\providecommand \@ifnum [1]{%
 \ifnum #1\expandafter \@firstoftwo
 \else \expandafter \@secondoftwo
 \fi
}%
\providecommand \@ifx [1]{%
 \ifx #1\expandafter \@firstoftwo
 \else \expandafter \@secondoftwo
 \fi
}%
\providecommand \natexlab [1]{#1}%
\providecommand \enquote  [1]{``#1''}%
\providecommand \bibnamefont  [1]{#1}%
\providecommand \bibfnamefont [1]{#1}%
\providecommand \citenamefont [1]{#1}%
\providecommand \href@noop [0]{\@secondoftwo}%
\providecommand \href [0]{\begingroup \@sanitize@url \@href}%
\providecommand \@href[1]{\@@startlink{#1}\@@href}%
\providecommand \@@href[1]{\endgroup#1\@@endlink}%
\providecommand \@sanitize@url [0]{\catcode `\\12\catcode `\$12\catcode
  `\&12\catcode `\#12\catcode `\^12\catcode `\_12\catcode `\%12\relax}%
\providecommand \@@startlink[1]{}%
\providecommand \@@endlink[0]{}%
\providecommand \url  [0]{\begingroup\@sanitize@url \@url }%
\providecommand \@url [1]{\endgroup\@href {#1}{\urlprefix }}%
\providecommand \urlprefix  [0]{URL }%
\providecommand \Eprint [0]{\href }%
\providecommand \doibase [0]{http://dx.doi.org/}%
\providecommand \selectlanguage [0]{\@gobble}%
\providecommand \bibinfo  [0]{\@secondoftwo}%
\providecommand \bibfield  [0]{\@secondoftwo}%
\providecommand \translation [1]{[#1]}%
\providecommand \BibitemOpen [0]{}%
\providecommand \bibitemStop [0]{}%
\providecommand \bibitemNoStop [0]{.\EOS\space}%
\providecommand \EOS [0]{\spacefactor3000\relax}%
\providecommand \BibitemShut  [1]{\csname bibitem#1\endcsname}%
\let\auto@bib@innerbib\@empty
%</preamble>
\bibitem [{\citenamefont {Graf}\ \emph {et~al.}(2011)\citenamefont {Graf},
  \citenamefont {Jozwiak}, \citenamefont {Smallwood}, \citenamefont {Eisaki},
  \citenamefont {Kaindl}, \citenamefont {Lee},\ and\ \citenamefont
  {Lanzara}}]{Graf2011}%
  \BibitemOpen
  \bibfield  {author} {\bibinfo {author} {\bibfnamefont {J.}~\bibnamefont
  {Graf}}, \bibinfo {author} {\bibfnamefont {C.}~\bibnamefont {Jozwiak}},
  \bibinfo {author} {\bibfnamefont {C.~L.}\ \bibnamefont {Smallwood}}, \bibinfo
  {author} {\bibfnamefont {H.}~\bibnamefont {Eisaki}}, \bibinfo {author}
  {\bibfnamefont {R.~A.}\ \bibnamefont {Kaindl}}, \bibinfo {author}
  {\bibfnamefont {D.-H.}\ \bibnamefont {Lee}}, \ and\ \bibinfo {author}
  {\bibfnamefont {A.}~\bibnamefont {Lanzara}},\ }\href {\doibase
  10.1038/nphys2027} {\bibfield  {journal} {\bibinfo  {journal} {Nature
  Physics}\ }\textbf {\bibinfo {volume} {7}},\ \bibinfo {pages} {805} (\bibinfo
  {year} {2011})}\BibitemShut {NoStop}%
\bibitem [{\citenamefont {Smallwood}\ \emph {et~al.}(2014)\citenamefont
  {Smallwood}, \citenamefont {Zhang}, \citenamefont {Miller}, \citenamefont
  {Jozwiak}, \citenamefont {Eisaki}, \citenamefont {Lee},\ and\ \citenamefont
  {Lanzara}}]{Smallwood2014}%
  \BibitemOpen
  \bibfield  {author} {\bibinfo {author} {\bibfnamefont {C.~L.}\ \bibnamefont
  {Smallwood}}, \bibinfo {author} {\bibfnamefont {W.}~\bibnamefont {Zhang}},
  \bibinfo {author} {\bibfnamefont {T.~L.}\ \bibnamefont {Miller}}, \bibinfo
  {author} {\bibfnamefont {C.}~\bibnamefont {Jozwiak}}, \bibinfo {author}
  {\bibfnamefont {H.}~\bibnamefont {Eisaki}}, \bibinfo {author} {\bibfnamefont
  {D.-H.}\ \bibnamefont {Lee}}, \ and\ \bibinfo {author} {\bibfnamefont
  {A.}~\bibnamefont {Lanzara}},\ }\href {\doibase 10.1103/PhysRevB.89.115126}
  {\bibfield  {journal} {\bibinfo  {journal} {Phys. Rev. B}\ }\textbf {\bibinfo
  {volume} {89}},\ \bibinfo {pages} {115126} (\bibinfo {year}
  {2014})}\BibitemShut {NoStop}%
\bibitem [{\citenamefont {Peronaci}\ \emph {et~al.}(2015)\citenamefont
  {Peronaci}, \citenamefont {Schir\'o},\ and\ \citenamefont
  {Capone}}]{Peronaci2015}%
  \BibitemOpen
  \bibfield  {author} {\bibinfo {author} {\bibfnamefont {F.}~\bibnamefont
  {Peronaci}}, \bibinfo {author} {\bibfnamefont {M.}~\bibnamefont {Schir\'o}},
  \ and\ \bibinfo {author} {\bibfnamefont {M.}~\bibnamefont {Capone}},\ }\href
  {\doibase 10.1103/PhysRevLett.115.257001} {\bibfield  {journal} {\bibinfo
  {journal} {Phys. Rev. Lett.}\ }\textbf {\bibinfo {volume} {115}},\ \bibinfo
  {pages} {257001} (\bibinfo {year} {2015})}\BibitemShut {NoStop}%
\bibitem [{\citenamefont {Caviglia}\ \emph {et~al.}(2012)\citenamefont
  {Caviglia}, \citenamefont {Scherwitzl}, \citenamefont {Popovich},
  \citenamefont {Hu}, \citenamefont {Bromberger}, \citenamefont {Singla},
  \citenamefont {Mitrano}, \citenamefont {Hoffmann}, \citenamefont {Kaiser},
  \citenamefont {Zubko}, \citenamefont {Gariglio}, \citenamefont {Triscone},
  \citenamefont {F\"orst},\ and\ \citenamefont {Cavalleri}}]{Caviglia2012}%
  \BibitemOpen
  \bibfield  {author} {\bibinfo {author} {\bibfnamefont {A.~D.}\ \bibnamefont
  {Caviglia}}, \bibinfo {author} {\bibfnamefont {R.}~\bibnamefont
  {Scherwitzl}}, \bibinfo {author} {\bibfnamefont {P.}~\bibnamefont
  {Popovich}}, \bibinfo {author} {\bibfnamefont {W.}~\bibnamefont {Hu}},
  \bibinfo {author} {\bibfnamefont {H.}~\bibnamefont {Bromberger}}, \bibinfo
  {author} {\bibfnamefont {R.}~\bibnamefont {Singla}}, \bibinfo {author}
  {\bibfnamefont {M.}~\bibnamefont {Mitrano}}, \bibinfo {author} {\bibfnamefont
  {M.~C.}\ \bibnamefont {Hoffmann}}, \bibinfo {author} {\bibfnamefont
  {S.}~\bibnamefont {Kaiser}}, \bibinfo {author} {\bibfnamefont
  {P.}~\bibnamefont {Zubko}}, \bibinfo {author} {\bibfnamefont
  {S.}~\bibnamefont {Gariglio}}, \bibinfo {author} {\bibfnamefont {J.-M.}\
  \bibnamefont {Triscone}}, \bibinfo {author} {\bibfnamefont {M.}~\bibnamefont
  {F\"orst}}, \ and\ \bibinfo {author} {\bibfnamefont {A.}~\bibnamefont
  {Cavalleri}},\ }\href {\doibase 10.1103/PhysRevLett.108.136801} {\bibfield
  {journal} {\bibinfo  {journal} {Phys. Rev. Lett.}\ }\textbf {\bibinfo
  {volume} {108}},\ \bibinfo {pages} {136801} (\bibinfo {year}
  {2012})}\BibitemShut {NoStop}%
\bibitem [{\citenamefont {Nava}\ \emph {et~al.}(2018)\citenamefont {Nava},
  \citenamefont {Giannetti}, \citenamefont {Georges}, \citenamefont {Tosatti},\
  and\ \citenamefont {Fabrizio}}]{Nava2018}%
  \BibitemOpen
  \bibfield  {author} {\bibinfo {author} {\bibfnamefont {A.}~\bibnamefont
  {Nava}}, \bibinfo {author} {\bibfnamefont {C.}~\bibnamefont {Giannetti}},
  \bibinfo {author} {\bibfnamefont {A.}~\bibnamefont {Georges}}, \bibinfo
  {author} {\bibfnamefont {E.}~\bibnamefont {Tosatti}}, \ and\ \bibinfo
  {author} {\bibfnamefont {M.}~\bibnamefont {Fabrizio}},\ }\href {\doibase
  10.1038/nphys4288} {\bibfield  {journal} {\bibinfo  {journal} {Nature
  Physics}\ }\textbf {\bibinfo {volume} {14}},\ \bibinfo {pages} {154}
  (\bibinfo {year} {2018})}\BibitemShut {NoStop}%
\bibitem [{\citenamefont {Lee}\ \emph {et~al.}(2006)\citenamefont {Lee},
  \citenamefont {Nagaosa},\ and\ \citenamefont {Wen}}]{Lee2006}%
  \BibitemOpen
  \bibfield  {author} {\bibinfo {author} {\bibfnamefont {P.~A.}\ \bibnamefont
  {Lee}}, \bibinfo {author} {\bibfnamefont {N.}~\bibnamefont {Nagaosa}}, \ and\
  \bibinfo {author} {\bibfnamefont {X.-G.}\ \bibnamefont {Wen}},\ }\href
  {\doibase 10.1103/RevModPhys.78.17} {\bibfield  {journal} {\bibinfo
  {journal} {Rev. Mod. Phys.}\ }\textbf {\bibinfo {volume} {78}},\ \bibinfo
  {pages} {17} (\bibinfo {year} {2006})}\BibitemShut {NoStop}%
\bibitem [{\citenamefont {Andr\'e}\ \emph {et~al.}(2012)\citenamefont
  {Andr\'e}, \citenamefont {Schir\'o},\ and\ \citenamefont
  {Fabrizio}}]{Andre2012}%
  \BibitemOpen
  \bibfield  {author} {\bibinfo {author} {\bibfnamefont {P.}~\bibnamefont
  {Andr\'e}}, \bibinfo {author} {\bibfnamefont {M.}~\bibnamefont {Schir\'o}}, \
  and\ \bibinfo {author} {\bibfnamefont {M.}~\bibnamefont {Fabrizio}},\ }\href
  {\doibase 10.1103/PhysRevB.85.205118} {\bibfield  {journal} {\bibinfo
  {journal} {Phys. Rev. B}\ }\textbf {\bibinfo {volume} {85}},\ \bibinfo
  {pages} {205118} (\bibinfo {year} {2012})}\BibitemShut {NoStop}%
\bibitem [{\citenamefont {Sandri}\ and\ \citenamefont
  {Fabrizio}(2015)}]{Sandri2015}%
  \BibitemOpen
  \bibfield  {author} {\bibinfo {author} {\bibfnamefont {M.}~\bibnamefont
  {Sandri}}\ and\ \bibinfo {author} {\bibfnamefont {M.}~\bibnamefont
  {Fabrizio}},\ }\href {\doibase 10.1103/PhysRevB.91.115102} {\bibfield
  {journal} {\bibinfo  {journal} {Phys. Rev. B}\ }\textbf {\bibinfo {volume}
  {91}},\ \bibinfo {pages} {115102} (\bibinfo {year} {2015})}\BibitemShut
  {NoStop}%
\bibitem [{\citenamefont {Fu}\ \emph {et~al.}(2014)\citenamefont {Fu},
  \citenamefont {Hung},\ and\ \citenamefont {Sachdev}}]{Fu2014}%
  \BibitemOpen
  \bibfield  {author} {\bibinfo {author} {\bibfnamefont {W.}~\bibnamefont
  {Fu}}, \bibinfo {author} {\bibfnamefont {L.-Y.}\ \bibnamefont {Hung}}, \ and\
  \bibinfo {author} {\bibfnamefont {S.}~\bibnamefont {Sachdev}},\ }\href
  {\doibase 10.1103/PhysRevB.90.024506} {\bibfield  {journal} {\bibinfo
  {journal} {Phys. Rev. B}\ }\textbf {\bibinfo {volume} {90}},\ \bibinfo
  {pages} {024506} (\bibinfo {year} {2014})}\BibitemShut {NoStop}%
\bibitem [{\citenamefont {Zvyagin}(2016)}]{Zvyagin2016}%
  \BibitemOpen
  \bibfield  {author} {\bibinfo {author} {\bibfnamefont {A.~A.}\ \bibnamefont
  {Zvyagin}},\ }\href {\doibase 10.1063/1.4969869} {\bibfield  {journal}
  {\bibinfo  {journal} {Low Temperature Physics}\ }\textbf {\bibinfo {volume}
  {42}},\ \bibinfo {pages} {971} (\bibinfo {year} {2016})}\BibitemShut
  {NoStop}%
\bibitem [{\citenamefont {Heyl}(2018)}]{Heyl2018}%
  \BibitemOpen
  \bibfield  {author} {\bibinfo {author} {\bibfnamefont {M.}~\bibnamefont
  {Heyl}},\ }\href {\doibase 10.1088/1361-6633/aaaf9a} {\bibfield  {journal}
  {\bibinfo  {journal} {Reports on Progress in Physics}\ }\textbf {\bibinfo
  {volume} {81}},\ \bibinfo {pages} {054001} (\bibinfo {year}
  {2018})}\BibitemShut {NoStop}%
\bibitem [{\citenamefont {Heyl}(2019)}]{Heyl2019}%
  \BibitemOpen
  \bibfield  {author} {\bibinfo {author} {\bibfnamefont {M.}~\bibnamefont
  {Heyl}},\ }\href {\doibase 10.1209/0295-5075/125/26001} {\bibfield  {journal}
  {\bibinfo  {journal} {Europhysics Letters}\ }\textbf {\bibinfo {volume}
  {125}},\ \bibinfo {pages} {26001} (\bibinfo {year} {2019})}\BibitemShut
  {NoStop}%
\bibitem [{\citenamefont {Mazza}(2017)}]{Mazza2017}%
  \BibitemOpen
  \bibfield  {author} {\bibinfo {author} {\bibfnamefont {G.}~\bibnamefont
  {Mazza}},\ }\href {\doibase 10.1103/PhysRevB.96.205110} {\bibfield  {journal}
  {\bibinfo  {journal} {Phys. Rev. B}\ }\textbf {\bibinfo {volume} {96}},\
  \bibinfo {pages} {205110} (\bibinfo {year} {2017})}\BibitemShut {NoStop}%
\bibitem [{\citenamefont {Breuer}\ and\ \citenamefont
  {Petruccione}(2002)}]{Petruccione2002}%
  \BibitemOpen
  \bibfield  {author} {\bibinfo {author} {\bibfnamefont {H.-P.}\ \bibnamefont
  {Breuer}}\ and\ \bibinfo {author} {\bibfnamefont {F.}~\bibnamefont
  {Petruccione}},\ }\href@noop {} {\emph {\bibinfo {title} {The Theory of Open
  Quantum Systems}}}\ (\bibinfo  {publisher} {Oxford University Press},\
  \bibinfo {year} {2002})\BibitemShut {NoStop}%
\bibitem [{\citenamefont {Wilde}(2013)}]{Wilde2013}%
  \BibitemOpen
  \bibfield  {author} {\bibinfo {author} {\bibfnamefont {M.~M.}\ \bibnamefont
  {Wilde}},\ }\href {\doibase 10.1017/CBO9781139525343} {\emph {\bibinfo
  {title} {Quantum Information Theory}}}\ (\bibinfo  {publisher} {Cambridge
  University Press},\ \bibinfo {year} {2013})\BibitemShut {NoStop}%
\bibitem [{\citenamefont {Nava}\ and\ \citenamefont
  {Fabrizio}(2019)}]{Nava2019}%
  \BibitemOpen
  \bibfield  {author} {\bibinfo {author} {\bibfnamefont {A.}~\bibnamefont
  {Nava}}\ and\ \bibinfo {author} {\bibfnamefont {M.}~\bibnamefont
  {Fabrizio}},\ }\href {\doibase 10.1103/PhysRevB.100.125102} {\bibfield
  {journal} {\bibinfo  {journal} {Phys. Rev. B}\ }\textbf {\bibinfo {volume}
  {100}},\ \bibinfo {pages} {125102} (\bibinfo {year} {2019})}\BibitemShut
  {NoStop}%
\bibitem [{\citenamefont {Manzano}(2020)}]{Manzano2020}%
  \BibitemOpen
  \bibfield  {author} {\bibinfo {author} {\bibfnamefont {D.}~\bibnamefont
  {Manzano}},\ }\href {https://doi.org/10.1063/1.5115323} {\bibfield  {journal}
  {\bibinfo  {journal} {AIP Advances}\ }\textbf {\bibinfo {volume} {10}}
  (\bibinfo {year} {2020})},\ \bibinfo {note} {025106}\BibitemShut {NoStop}%
\bibitem [{\citenamefont {Nava}\ and\ \citenamefont
  {Fabrizio}(2022)}]{Nava2022}%
  \BibitemOpen
  \bibfield  {author} {\bibinfo {author} {\bibfnamefont {A.}~\bibnamefont
  {Nava}}\ and\ \bibinfo {author} {\bibfnamefont {M.}~\bibnamefont
  {Fabrizio}},\ }\href {\doibase 10.21468/SciPostPhys.12.1.014} {\bibfield
  {journal} {\bibinfo  {journal} {SciPost Phys.}\ }\textbf {\bibinfo {volume}
  {12}},\ \bibinfo {pages} {014} (\bibinfo {year} {2022})}\BibitemShut
  {NoStop}%
\bibitem [{\citenamefont {Artiaco}\ \emph {et~al.}(2023)\citenamefont
  {Artiaco}, \citenamefont {Nava},\ and\ \citenamefont
  {Fabrizio}}]{Artiaco2023}%
  \BibitemOpen
  \bibfield  {author} {\bibinfo {author} {\bibfnamefont {C.}~\bibnamefont
  {Artiaco}}, \bibinfo {author} {\bibfnamefont {A.}~\bibnamefont {Nava}}, \
  and\ \bibinfo {author} {\bibfnamefont {M.}~\bibnamefont {Fabrizio}},\ }\href
  {\doibase 10.1103/PhysRevB.107.104201} {\bibfield  {journal} {\bibinfo
  {journal} {Phys. Rev. B}\ }\textbf {\bibinfo {volume} {107}},\ \bibinfo
  {pages} {104201} (\bibinfo {year} {2023})}\BibitemShut {NoStop}%
\bibitem [{\citenamefont {Mazza}\ and\ \citenamefont
  {Schir\`o}(2023)}]{Mazza2023}%
  \BibitemOpen
  \bibfield  {author} {\bibinfo {author} {\bibfnamefont {G.}~\bibnamefont
  {Mazza}}\ and\ \bibinfo {author} {\bibfnamefont {M.}~\bibnamefont
  {Schir\`o}},\ }\href {\doibase 10.1103/PhysRevA.107.L051301} {\bibfield
  {journal} {\bibinfo  {journal} {Phys. Rev. A}\ }\textbf {\bibinfo {volume}
  {107}},\ \bibinfo {pages} {L051301} (\bibinfo {year} {2023})}\BibitemShut
  {NoStop}%
\bibitem [{\citenamefont {Cui}\ \emph {et~al.}(2019)\citenamefont {Cui},
  \citenamefont {Yang}, \citenamefont {Vaswani}, \citenamefont {Wang},
  \citenamefont {Fernandes},\ and\ \citenamefont {Orth}}]{Cui2019}%
  \BibitemOpen
  \bibfield  {author} {\bibinfo {author} {\bibfnamefont {T.}~\bibnamefont
  {Cui}}, \bibinfo {author} {\bibfnamefont {X.}~\bibnamefont {Yang}}, \bibinfo
  {author} {\bibfnamefont {C.}~\bibnamefont {Vaswani}}, \bibinfo {author}
  {\bibfnamefont {J.}~\bibnamefont {Wang}}, \bibinfo {author} {\bibfnamefont
  {R.~M.}\ \bibnamefont {Fernandes}}, \ and\ \bibinfo {author} {\bibfnamefont
  {P.~P.}\ \bibnamefont {Orth}},\ }\href {\doibase 10.1103/PhysRevB.100.054504}
  {\bibfield  {journal} {\bibinfo  {journal} {Phys. Rev. B}\ }\textbf {\bibinfo
  {volume} {100}},\ \bibinfo {pages} {054504} (\bibinfo {year}
  {2019})}\BibitemShut {NoStop}%
\bibitem [{\citenamefont {Heyl}\ \emph {et~al.}(2013)\citenamefont {Heyl},
  \citenamefont {Polkovnikov},\ and\ \citenamefont {Kehrein}}]{Heyl2013}%
  \BibitemOpen
  \bibfield  {author} {\bibinfo {author} {\bibfnamefont {M.}~\bibnamefont
  {Heyl}}, \bibinfo {author} {\bibfnamefont {A.}~\bibnamefont {Polkovnikov}}, \
  and\ \bibinfo {author} {\bibfnamefont {S.}~\bibnamefont {Kehrein}},\ }\href
  {\doibase 10.1103/PhysRevLett.110.135704} {\bibfield  {journal} {\bibinfo
  {journal} {Phys. Rev. Lett.}\ }\textbf {\bibinfo {volume} {110}},\ \bibinfo
  {pages} {135704} (\bibinfo {year} {2013})}\BibitemShut {NoStop}%
\bibitem [{\citenamefont {Jurcevic}\ \emph {et~al.}(2017)\citenamefont
  {Jurcevic}, \citenamefont {Shen}, \citenamefont {Hauke}, \citenamefont
  {Maier}, \citenamefont {Brydges}, \citenamefont {Hempel}, \citenamefont
  {Lanyon}, \citenamefont {Heyl}, \citenamefont {Blatt},\ and\ \citenamefont
  {Roos}}]{Jurcevic2017}%
  \BibitemOpen
  \bibfield  {author} {\bibinfo {author} {\bibfnamefont {P.}~\bibnamefont
  {Jurcevic}}, \bibinfo {author} {\bibfnamefont {H.}~\bibnamefont {Shen}},
  \bibinfo {author} {\bibfnamefont {P.}~\bibnamefont {Hauke}}, \bibinfo
  {author} {\bibfnamefont {C.}~\bibnamefont {Maier}}, \bibinfo {author}
  {\bibfnamefont {T.}~\bibnamefont {Brydges}}, \bibinfo {author} {\bibfnamefont
  {C.}~\bibnamefont {Hempel}}, \bibinfo {author} {\bibfnamefont {B.~P.}\
  \bibnamefont {Lanyon}}, \bibinfo {author} {\bibfnamefont {M.}~\bibnamefont
  {Heyl}}, \bibinfo {author} {\bibfnamefont {R.}~\bibnamefont {Blatt}}, \ and\
  \bibinfo {author} {\bibfnamefont {C.~F.}\ \bibnamefont {Roos}},\ }\href
  {\doibase 10.1103/PhysRevLett.119.080501} {\bibfield  {journal} {\bibinfo
  {journal} {Phys. Rev. Lett.}\ }\textbf {\bibinfo {volume} {119}},\ \bibinfo
  {pages} {080501} (\bibinfo {year} {2017})}\BibitemShut {NoStop}%
\bibitem [{\citenamefont {Schmied}\ \emph {et~al.}(2019)\citenamefont
  {Schmied}, \citenamefont {Mikheev},\ and\ \citenamefont
  {Gasenzer}}]{Schmied2019}%
  \BibitemOpen
  \bibfield  {author} {\bibinfo {author} {\bibfnamefont {C.-M.}\ \bibnamefont
  {Schmied}}, \bibinfo {author} {\bibfnamefont {A.~N.}\ \bibnamefont
  {Mikheev}}, \ and\ \bibinfo {author} {\bibfnamefont {T.}~\bibnamefont
  {Gasenzer}},\ }\href {\doibase 10.1142/S0217751X19410069} {\bibfield
  {journal} {\bibinfo  {journal} {International Journal of Modern Physics A}\
  }\textbf {\bibinfo {volume} {34}},\ \bibinfo {pages} {1941006} (\bibinfo
  {year} {2019})}\BibitemShut {NoStop}%
\bibitem [{\citenamefont {Yuzbashyan}\ and\ \citenamefont
  {Dzero}(2006)}]{Yuzbashyan2006b}%
  \BibitemOpen
  \bibfield  {author} {\bibinfo {author} {\bibfnamefont {E.~A.}\ \bibnamefont
  {Yuzbashyan}}\ and\ \bibinfo {author} {\bibfnamefont {M.}~\bibnamefont
  {Dzero}},\ }\href {\doibase 10.1103/PhysRevLett.96.230404} {\bibfield
  {journal} {\bibinfo  {journal} {Phys. Rev. Lett.}\ }\textbf {\bibinfo
  {volume} {96}},\ \bibinfo {pages} {230404} (\bibinfo {year}
  {2006})}\BibitemShut {NoStop}%
\bibitem [{\citenamefont {Pr{\"u}fer}\ \emph {et~al.}(2018)\citenamefont
  {Pr{\"u}fer}, \citenamefont {Kunkel}, \citenamefont {Strobel}, \citenamefont
  {Lannig}, \citenamefont {Linnemann}, \citenamefont {Schmied}, \citenamefont
  {Berges}, \citenamefont {Gasenzer},\ and\ \citenamefont
  {Oberthaler}}]{Prufer2018}%
  \BibitemOpen
  \bibfield  {author} {\bibinfo {author} {\bibfnamefont {M.}~\bibnamefont
  {Pr{\"u}fer}}, \bibinfo {author} {\bibfnamefont {P.}~\bibnamefont {Kunkel}},
  \bibinfo {author} {\bibfnamefont {H.}~\bibnamefont {Strobel}}, \bibinfo
  {author} {\bibfnamefont {S.}~\bibnamefont {Lannig}}, \bibinfo {author}
  {\bibfnamefont {D.}~\bibnamefont {Linnemann}}, \bibinfo {author}
  {\bibfnamefont {C.-M.}\ \bibnamefont {Schmied}}, \bibinfo {author}
  {\bibfnamefont {J.}~\bibnamefont {Berges}}, \bibinfo {author} {\bibfnamefont
  {T.}~\bibnamefont {Gasenzer}}, \ and\ \bibinfo {author} {\bibfnamefont
  {M.~K.}\ \bibnamefont {Oberthaler}},\ }\href {\doibase
  10.1038/s41586-018-0659-0} {\bibfield  {journal} {\bibinfo  {journal}
  {Nature}\ }\textbf {\bibinfo {volume} {563}},\ \bibinfo {pages} {217}
  (\bibinfo {year} {2018})}\BibitemShut {NoStop}%
\bibitem [{\citenamefont {Yamamoto}\ \emph {et~al.}(2021)\citenamefont
  {Yamamoto}, \citenamefont {Nakagawa}, \citenamefont {Tsuji}, \citenamefont
  {Ueda},\ and\ \citenamefont {Kawakami}}]{Yamamoto2021}%
  \BibitemOpen
  \bibfield  {author} {\bibinfo {author} {\bibfnamefont {K.}~\bibnamefont
  {Yamamoto}}, \bibinfo {author} {\bibfnamefont {M.}~\bibnamefont {Nakagawa}},
  \bibinfo {author} {\bibfnamefont {N.}~\bibnamefont {Tsuji}}, \bibinfo
  {author} {\bibfnamefont {M.}~\bibnamefont {Ueda}}, \ and\ \bibinfo {author}
  {\bibfnamefont {N.}~\bibnamefont {Kawakami}},\ }\href {\doibase
  10.1103/PhysRevLett.127.055301} {\bibfield  {journal} {\bibinfo  {journal}
  {Phys. Rev. Lett.}\ }\textbf {\bibinfo {volume} {127}},\ \bibinfo {pages}
  {055301} (\bibinfo {year} {2021})}\BibitemShut {NoStop}%
\bibitem [{\citenamefont {Mondal}\ and\ \citenamefont
  {Nag}(2022)}]{Debashish2022}%
  \BibitemOpen
  \bibfield  {author} {\bibinfo {author} {\bibfnamefont {D.}~\bibnamefont
  {Mondal}}\ and\ \bibinfo {author} {\bibfnamefont {T.}~\bibnamefont {Nag}},\
  }\href {\doibase 10.1103/PhysRevB.106.054308} {\bibfield  {journal} {\bibinfo
   {journal} {Phys. Rev. B}\ }\textbf {\bibinfo {volume} {106}},\ \bibinfo
  {pages} {054308} (\bibinfo {year} {2022})}\BibitemShut {NoStop}%
\bibitem [{\citenamefont {Mondal}\ and\ \citenamefont
  {Nag}(2023)}]{Debashish2023}%
  \BibitemOpen
  \bibfield  {author} {\bibinfo {author} {\bibfnamefont {D.}~\bibnamefont
  {Mondal}}\ and\ \bibinfo {author} {\bibfnamefont {T.}~\bibnamefont {Nag}},\
  }\href {\doibase 10.1103/PhysRevB.107.184311} {\bibfield  {journal} {\bibinfo
   {journal} {Phys. Rev. B}\ }\textbf {\bibinfo {volume} {107}},\ \bibinfo
  {pages} {184311} (\bibinfo {year} {2023})}\BibitemShut {NoStop}%
\bibitem [{\citenamefont {Pollmann}\ \emph {et~al.}(2010)\citenamefont
  {Pollmann}, \citenamefont {Mukerjee}, \citenamefont {Green},\ and\
  \citenamefont {Moore}}]{Pollmann2010}%
  \BibitemOpen
  \bibfield  {author} {\bibinfo {author} {\bibfnamefont {F.}~\bibnamefont
  {Pollmann}}, \bibinfo {author} {\bibfnamefont {S.}~\bibnamefont {Mukerjee}},
  \bibinfo {author} {\bibfnamefont {A.~G.}\ \bibnamefont {Green}}, \ and\
  \bibinfo {author} {\bibfnamefont {J.~E.}\ \bibnamefont {Moore}},\ }\href
  {\doibase 10.1103/PhysRevE.81.020101} {\bibfield  {journal} {\bibinfo
  {journal} {Phys. Rev. E}\ }\textbf {\bibinfo {volume} {81}},\ \bibinfo
  {pages} {020101} (\bibinfo {year} {2010})}\BibitemShut {NoStop}%
\bibitem [{\citenamefont {Abeling}\ and\ \citenamefont
  {Kehrein}(2016)}]{Abeling2016}%
  \BibitemOpen
  \bibfield  {author} {\bibinfo {author} {\bibfnamefont {N.~O.}\ \bibnamefont
  {Abeling}}\ and\ \bibinfo {author} {\bibfnamefont {S.}~\bibnamefont
  {Kehrein}},\ }\href {\doibase 10.1103/PhysRevB.93.104302} {\bibfield
  {journal} {\bibinfo  {journal} {Phys. Rev. B}\ }\textbf {\bibinfo {volume}
  {93}},\ \bibinfo {pages} {104302} (\bibinfo {year} {2016})}\BibitemShut
  {NoStop}%
\bibitem [{\citenamefont {Bhattacharya}\ \emph {et~al.}(2017)\citenamefont
  {Bhattacharya}, \citenamefont {Bandyopadhyay},\ and\ \citenamefont
  {Dutta}}]{Bhattacharya2017}%
  \BibitemOpen
  \bibfield  {author} {\bibinfo {author} {\bibfnamefont {U.}~\bibnamefont
  {Bhattacharya}}, \bibinfo {author} {\bibfnamefont {S.}~\bibnamefont
  {Bandyopadhyay}}, \ and\ \bibinfo {author} {\bibfnamefont {A.}~\bibnamefont
  {Dutta}},\ }\href {\doibase 10.1103/PhysRevB.96.180303} {\bibfield  {journal}
  {\bibinfo  {journal} {Phys. Rev. B}\ }\textbf {\bibinfo {volume} {96}},\
  \bibinfo {pages} {180303} (\bibinfo {year} {2017})}\BibitemShut {NoStop}%
\bibitem [{\citenamefont {Lang}\ \emph {et~al.}(2018)\citenamefont {Lang},
  \citenamefont {Frank},\ and\ \citenamefont {Halimeh}}]{Lang2018}%
  \BibitemOpen
  \bibfield  {author} {\bibinfo {author} {\bibfnamefont {J.}~\bibnamefont
  {Lang}}, \bibinfo {author} {\bibfnamefont {B.}~\bibnamefont {Frank}}, \ and\
  \bibinfo {author} {\bibfnamefont {J.~C.}\ \bibnamefont {Halimeh}},\ }\href
  {\doibase 10.1103/PhysRevB.97.174401} {\bibfield  {journal} {\bibinfo
  {journal} {Phys. Rev. B}\ }\textbf {\bibinfo {volume} {97}},\ \bibinfo
  {pages} {174401} (\bibinfo {year} {2018})}\BibitemShut {NoStop}%
\bibitem [{\citenamefont {Wu}\ \emph {et~al.}(2022)\citenamefont {Wu},
  \citenamefont {Nettersheim}, \citenamefont {Fe\ss}, \citenamefont {Schnell},
  \citenamefont {Burgardt}, \citenamefont {Hiebel}, \citenamefont {Adam},
  \citenamefont {Eckardt},\ and\ \citenamefont {Widera}}]{Wu2022}%
  \BibitemOpen
  \bibfield  {author} {\bibinfo {author} {\bibfnamefont {L.-N.}\ \bibnamefont
  {Wu}}, \bibinfo {author} {\bibfnamefont {J.}~\bibnamefont {Nettersheim}},
  \bibinfo {author} {\bibfnamefont {J.}~\bibnamefont {Fe\ss}}, \bibinfo
  {author} {\bibfnamefont {A.}~\bibnamefont {Schnell}}, \bibinfo {author}
  {\bibfnamefont {S.}~\bibnamefont {Burgardt}}, \bibinfo {author}
  {\bibfnamefont {S.}~\bibnamefont {Hiebel}}, \bibinfo {author} {\bibfnamefont
  {D.}~\bibnamefont {Adam}}, \bibinfo {author} {\bibfnamefont {A.}~\bibnamefont
  {Eckardt}}, \ and\ \bibinfo {author} {\bibfnamefont {A.}~\bibnamefont
  {Widera}},\ }\href@noop {} {\enquote {\bibinfo {title} {Dynamical phase
  transition in an open quantum system},}\ } (\bibinfo {year} {2022}),\ \Eprint
  {http://arxiv.org/abs/2208.05164} {arXiv:2208.05164 [cond-mat.quant-gas]}
  \BibitemShut {NoStop}%
\bibitem [{\citenamefont {Biscaras}\ \emph {et~al.}(2012)\citenamefont
  {Biscaras}, \citenamefont {Bergeal}, \citenamefont {Hurand}, \citenamefont
  {Grosset\^ete}, \citenamefont {Rastogi}, \citenamefont {Budhani},
  \citenamefont {LeBoeuf}, \citenamefont {Proust},\ and\ \citenamefont
  {Lesueur}}]{Biscaras2012}%
  \BibitemOpen
  \bibfield  {author} {\bibinfo {author} {\bibfnamefont {J.}~\bibnamefont
  {Biscaras}}, \bibinfo {author} {\bibfnamefont {N.}~\bibnamefont {Bergeal}},
  \bibinfo {author} {\bibfnamefont {S.}~\bibnamefont {Hurand}}, \bibinfo
  {author} {\bibfnamefont {C.}~\bibnamefont {Grosset\^ete}}, \bibinfo {author}
  {\bibfnamefont {A.}~\bibnamefont {Rastogi}}, \bibinfo {author} {\bibfnamefont
  {R.~C.}\ \bibnamefont {Budhani}}, \bibinfo {author} {\bibfnamefont
  {D.}~\bibnamefont {LeBoeuf}}, \bibinfo {author} {\bibfnamefont
  {C.}~\bibnamefont {Proust}}, \ and\ \bibinfo {author} {\bibfnamefont
  {J.}~\bibnamefont {Lesueur}},\ }\href {\doibase
  10.1103/PhysRevLett.108.247004} {\bibfield  {journal} {\bibinfo  {journal}
  {Phys. Rev. Lett.}\ }\textbf {\bibinfo {volume} {108}},\ \bibinfo {pages}
  {247004} (\bibinfo {year} {2012})}\BibitemShut {NoStop}%
\bibitem [{\citenamefont {Scheurer}\ and\ \citenamefont
  {Schmalian}(2015)}]{Scheurer2015}%
  \BibitemOpen
  \bibfield  {author} {\bibinfo {author} {\bibfnamefont {M.~S.}\ \bibnamefont
  {Scheurer}}\ and\ \bibinfo {author} {\bibfnamefont {J.}~\bibnamefont
  {Schmalian}},\ }\href {\doibase 10.1038/ncomms7005} {\bibfield  {journal}
  {\bibinfo  {journal} {Nature Communications}\ }\textbf {\bibinfo {volume}
  {6}},\ \bibinfo {pages} {6005} (\bibinfo {year} {2015})}\BibitemShut
  {NoStop}%
\bibitem [{\citenamefont {Perroni}\ \emph {et~al.}(2019)\citenamefont
  {Perroni}, \citenamefont {Cataudella}, \citenamefont {Salluzzo},
  \citenamefont {Cuoco},\ and\ \citenamefont {Citro}}]{Perroni2019}%
  \BibitemOpen
  \bibfield  {author} {\bibinfo {author} {\bibfnamefont {C.~A.}\ \bibnamefont
  {Perroni}}, \bibinfo {author} {\bibfnamefont {V.}~\bibnamefont {Cataudella}},
  \bibinfo {author} {\bibfnamefont {M.}~\bibnamefont {Salluzzo}}, \bibinfo
  {author} {\bibfnamefont {M.}~\bibnamefont {Cuoco}}, \ and\ \bibinfo {author}
  {\bibfnamefont {R.}~\bibnamefont {Citro}},\ }\href {\doibase
  10.1103/PhysRevB.100.094526} {\bibfield  {journal} {\bibinfo  {journal}
  {Phys. Rev. B}\ }\textbf {\bibinfo {volume} {100}},\ \bibinfo {pages}
  {094526} (\bibinfo {year} {2019})}\BibitemShut {NoStop}%
\bibitem [{\citenamefont {Lepori}\ \emph {et~al.}(2021)\citenamefont {Lepori},
  \citenamefont {Giuliano}, \citenamefont {Nava},\ and\ \citenamefont
  {Perroni}}]{Lepori2021}%
  \BibitemOpen
  \bibfield  {author} {\bibinfo {author} {\bibfnamefont {L.}~\bibnamefont
  {Lepori}}, \bibinfo {author} {\bibfnamefont {D.}~\bibnamefont {Giuliano}},
  \bibinfo {author} {\bibfnamefont {A.}~\bibnamefont {Nava}}, \ and\ \bibinfo
  {author} {\bibfnamefont {C.~A.}\ \bibnamefont {Perroni}},\ }\href {\doibase
  10.1103/PhysRevB.104.134509} {\bibfield  {journal} {\bibinfo  {journal}
  {Phys. Rev. B}\ }\textbf {\bibinfo {volume} {104}},\ \bibinfo {pages}
  {134509} (\bibinfo {year} {2021})}\BibitemShut {NoStop}%
\bibitem [{\citenamefont {Nava}\ \emph
  {et~al.}(2023{\natexlab{a}})\citenamefont {Nava}, \citenamefont {Perroni},
  \citenamefont {Egger}, \citenamefont {Lepori},\ and\ \citenamefont
  {Giuliano}}]{shorter_paper}%
  \BibitemOpen
  \bibfield  {author} {\bibinfo {author} {\bibfnamefont {A.}~\bibnamefont
  {Nava}}, \bibinfo {author} {\bibfnamefont {C.~A.}\ \bibnamefont {Perroni}},
  \bibinfo {author} {\bibfnamefont {R.}~\bibnamefont {Egger}}, \bibinfo
  {author} {\bibfnamefont {L.}~\bibnamefont {Lepori}}, \ and\ \bibinfo {author}
  {\bibfnamefont {D.}~\bibnamefont {Giuliano}},\ }\href@noop {} {\enquote
  {\bibinfo {title} {Dissipation driven dynamical topological phase transitions
  in two-dimensional superconductors},}\ } (\bibinfo {year}
  {2023}{\natexlab{a}}),\ \Eprint {http://arxiv.org/abs/2308.08265}
  {arXiv:2308.08265 [cond-mat.str-el]} \BibitemShut {NoStop}%
\bibitem [{\citenamefont {Laughlin}(1998)}]{Laughlin1998}%
  \BibitemOpen
  \bibfield  {author} {\bibinfo {author} {\bibfnamefont {R.~B.}\ \bibnamefont
  {Laughlin}},\ }\href {\doibase 10.1103/PhysRevLett.80.5188} {\bibfield
  {journal} {\bibinfo  {journal} {Phys. Rev. Lett.}\ }\textbf {\bibinfo
  {volume} {80}},\ \bibinfo {pages} {5188} (\bibinfo {year}
  {1998})}\BibitemShut {NoStop}%
\bibitem [{\citenamefont {Ghosh}\ and\ \citenamefont
  {Adhikari}(1999)}]{Ghosh1999}%
  \BibitemOpen
  \bibfield  {author} {\bibinfo {author} {\bibfnamefont {A.}~\bibnamefont
  {Ghosh}}\ and\ \bibinfo {author} {\bibfnamefont {S.~K.}\ \bibnamefont
  {Adhikari}},\ }\href {\doibase 10.1103/PhysRevB.60.10401} {\bibfield
  {journal} {\bibinfo  {journal} {Phys. Rev. B}\ }\textbf {\bibinfo {volume}
  {60}},\ \bibinfo {pages} {10401} (\bibinfo {year} {1999})}\BibitemShut
  {NoStop}%
\bibitem [{\citenamefont {Salkola}\ and\ \citenamefont
  {Schrieffer}(1998)}]{Salkola1998}%
  \BibitemOpen
  \bibfield  {author} {\bibinfo {author} {\bibfnamefont {M.~I.}\ \bibnamefont
  {Salkola}}\ and\ \bibinfo {author} {\bibfnamefont {J.~R.}\ \bibnamefont
  {Schrieffer}},\ }\href {\doibase 10.1103/PhysRevB.58.R5952} {\bibfield
  {journal} {\bibinfo  {journal} {Phys. Rev. B}\ }\textbf {\bibinfo {volume}
  {58}},\ \bibinfo {pages} {R5952} (\bibinfo {year} {1998})}\BibitemShut
  {NoStop}%
\bibitem [{\citenamefont {Ghosh}\ and\ \citenamefont
  {Adhikari}(2002)}]{Ghosh2002}%
  \BibitemOpen
  \bibfield  {author} {\bibinfo {author} {\bibfnamefont {A.}~\bibnamefont
  {Ghosh}}\ and\ \bibinfo {author} {\bibfnamefont {S.~K.}\ \bibnamefont
  {Adhikari}},\ }\href {\doibase https://doi.org/10.1016/S0921-4534(01)00932-7}
  {\bibfield  {journal} {\bibinfo  {journal} {Physica C: Superconductivity}\
  }\textbf {\bibinfo {volume} {370}},\ \bibinfo {pages} {146} (\bibinfo {year}
  {2002})}\BibitemShut {NoStop}%
\bibitem [{\citenamefont {Goldman}\ \emph {et~al.}(2016)\citenamefont
  {Goldman}, \citenamefont {Budich},\ and\ \citenamefont
  {Zoller}}]{Goldman2016}%
  \BibitemOpen
  \bibfield  {author} {\bibinfo {author} {\bibfnamefont {N.}~\bibnamefont
  {Goldman}}, \bibinfo {author} {\bibfnamefont {J.~C.}\ \bibnamefont {Budich}},
  \ and\ \bibinfo {author} {\bibfnamefont {P.}~\bibnamefont {Zoller}},\ }\href
  {\doibase 10.1038/nphys3803} {\bibfield  {journal} {\bibinfo  {journal}
  {Nature Physics}\ }\textbf {\bibinfo {volume} {12}},\ \bibinfo {pages} {639}
  (\bibinfo {year} {2016})}\BibitemShut {NoStop}%
\bibitem [{\citenamefont {Micnas}\ \emph {et~al.}(1990)\citenamefont {Micnas},
  \citenamefont {Ranninger},\ and\ \citenamefont {Robaszkiewicz}}]{Micnas1990}%
  \BibitemOpen
  \bibfield  {author} {\bibinfo {author} {\bibfnamefont {R.}~\bibnamefont
  {Micnas}}, \bibinfo {author} {\bibfnamefont {J.}~\bibnamefont {Ranninger}}, \
  and\ \bibinfo {author} {\bibfnamefont {S.}~\bibnamefont {Robaszkiewicz}},\
  }\href {\doibase 10.1103/RevModPhys.62.113} {\bibfield  {journal} {\bibinfo
  {journal} {Rev. Mod. Phys.}\ }\textbf {\bibinfo {volume} {62}},\ \bibinfo
  {pages} {113} (\bibinfo {year} {1990})}\BibitemShut {NoStop}%
\bibitem [{\citenamefont {Tsuei}\ and\ \citenamefont
  {Kirtley}(2000)}]{Tsuei2000}%
  \BibitemOpen
  \bibfield  {author} {\bibinfo {author} {\bibfnamefont {C.~C.}\ \bibnamefont
  {Tsuei}}\ and\ \bibinfo {author} {\bibfnamefont {J.~R.}\ \bibnamefont
  {Kirtley}},\ }\href {\doibase 10.1103/RevModPhys.72.969} {\bibfield
  {journal} {\bibinfo  {journal} {Rev. Mod. Phys.}\ }\textbf {\bibinfo {volume}
  {72}},\ \bibinfo {pages} {969} (\bibinfo {year} {2000})}\BibitemShut
  {NoStop}%
\bibitem [{\citenamefont {Balatsky}(1998)}]{Balatsky1998}%
  \BibitemOpen
  \bibfield  {author} {\bibinfo {author} {\bibfnamefont {A.~V.}\ \bibnamefont
  {Balatsky}},\ }\href {\doibase 10.1103/PhysRevLett.80.1972} {\bibfield
  {journal} {\bibinfo  {journal} {Phys. Rev. Lett.}\ }\textbf {\bibinfo
  {volume} {80}},\ \bibinfo {pages} {1972} (\bibinfo {year}
  {1998})}\BibitemShut {NoStop}%
\bibitem [{\citenamefont {Gor'kov}\ and\ \citenamefont
  {Rashba}(2001)}]{Gorkov2001}%
  \BibitemOpen
  \bibfield  {author} {\bibinfo {author} {\bibfnamefont {L.~P.}\ \bibnamefont
  {Gor'kov}}\ and\ \bibinfo {author} {\bibfnamefont {E.~I.}\ \bibnamefont
  {Rashba}},\ }\href {\doibase 10.1103/PhysRevLett.87.037004} {\bibfield
  {journal} {\bibinfo  {journal} {Phys. Rev. Lett.}\ }\textbf {\bibinfo
  {volume} {87}},\ \bibinfo {pages} {037004} (\bibinfo {year}
  {2001})}\BibitemShut {NoStop}%
\bibitem [{\citenamefont {Chern}(2016)}]{Chern2016}%
  \BibitemOpen
  \bibfield  {author} {\bibinfo {author} {\bibfnamefont {T.}~\bibnamefont
  {Chern}},\ }\href {\doibase 10.1063/1.4961462} {\bibfield  {journal}
  {\bibinfo  {journal} {AIP Advances}\ }\textbf {\bibinfo {volume} {6}},\
  \bibinfo {pages} {085211} (\bibinfo {year} {2016})}\BibitemShut {NoStop}%
\bibitem [{\citenamefont {Mitrano}\ \emph {et~al.}(2016)\citenamefont
  {Mitrano}, \citenamefont {Cantaluppi}, \citenamefont {Nicoletti},
  \citenamefont {Kaiser}, \citenamefont {Perucchi}, \citenamefont {Lupi},
  \citenamefont {Di~Pietro}, \citenamefont {Pontiroli}, \citenamefont
  {Ricc{\`o}}, \citenamefont {Clark}, \citenamefont {Jaksch},\ and\
  \citenamefont {Cavalleri}}]{Mitrano2016}%
  \BibitemOpen
  \bibfield  {author} {\bibinfo {author} {\bibfnamefont {M.}~\bibnamefont
  {Mitrano}}, \bibinfo {author} {\bibfnamefont {A.}~\bibnamefont {Cantaluppi}},
  \bibinfo {author} {\bibfnamefont {D.}~\bibnamefont {Nicoletti}}, \bibinfo
  {author} {\bibfnamefont {S.}~\bibnamefont {Kaiser}}, \bibinfo {author}
  {\bibfnamefont {A.}~\bibnamefont {Perucchi}}, \bibinfo {author}
  {\bibfnamefont {S.}~\bibnamefont {Lupi}}, \bibinfo {author} {\bibfnamefont
  {P.}~\bibnamefont {Di~Pietro}}, \bibinfo {author} {\bibfnamefont
  {D.}~\bibnamefont {Pontiroli}}, \bibinfo {author} {\bibfnamefont
  {M.}~\bibnamefont {Ricc{\`o}}}, \bibinfo {author} {\bibfnamefont {S.~R.}\
  \bibnamefont {Clark}}, \bibinfo {author} {\bibfnamefont {D.}~\bibnamefont
  {Jaksch}}, \ and\ \bibinfo {author} {\bibfnamefont {A.}~\bibnamefont
  {Cavalleri}},\ }\href {\doibase 10.1038/nature16522} {\bibfield  {journal}
  {\bibinfo  {journal} {Nature}\ }\textbf {\bibinfo {volume} {530}},\ \bibinfo
  {pages} {461} (\bibinfo {year} {2016})}\BibitemShut {NoStop}%
\bibitem [{\citenamefont {Choi}\ \emph {et~al.}(2023)\citenamefont {Choi},
  \citenamefont {Jeong}, \citenamefont {Min}, \citenamefont {Lee},
  \citenamefont {Choi},\ and\ \citenamefont {Lee}}]{Choi2023}%
  \BibitemOpen
  \bibfield  {author} {\bibinfo {author} {\bibfnamefont {I.~H.}\ \bibnamefont
  {Choi}}, \bibinfo {author} {\bibfnamefont {S.~G.}\ \bibnamefont {Jeong}},
  \bibinfo {author} {\bibfnamefont {T.}~\bibnamefont {Min}}, \bibinfo {author}
  {\bibfnamefont {J.}~\bibnamefont {Lee}}, \bibinfo {author} {\bibfnamefont
  {W.~S.}\ \bibnamefont {Choi}}, \ and\ \bibinfo {author} {\bibfnamefont
  {J.~S.}\ \bibnamefont {Lee}},\ }\href {\doibase
  https://doi.org/10.1002/advs.202300012} {\bibfield  {journal} {\bibinfo
  {journal} {Advanced Science}\ }\textbf {\bibinfo {volume} {10}},\ \bibinfo
  {pages} {2300012} (\bibinfo {year} {2023})},\ \Eprint
  {http://arxiv.org/abs/https://onlinelibrary.wiley.com/doi/pdf/10.1002/advs.202300012}
  {https://onlinelibrary.wiley.com/doi/pdf/10.1002/advs.202300012} \BibitemShut
  {NoStop}%
\bibitem [{\citenamefont {Huang}\ \emph {et~al.}(2023)\citenamefont {Huang},
  \citenamefont {Yue}, \citenamefont {Baydin}, \citenamefont {Zhu},
  \citenamefont {Nojiri}, \citenamefont {Kono}, \citenamefont {He},\ and\
  \citenamefont {Yi}}]{Jianwei2023}%
  \BibitemOpen
  \bibfield  {author} {\bibinfo {author} {\bibfnamefont {J.}~\bibnamefont
  {Huang}}, \bibinfo {author} {\bibfnamefont {Z.}~\bibnamefont {Yue}}, \bibinfo
  {author} {\bibfnamefont {A.}~\bibnamefont {Baydin}}, \bibinfo {author}
  {\bibfnamefont {H.}~\bibnamefont {Zhu}}, \bibinfo {author} {\bibfnamefont
  {H.}~\bibnamefont {Nojiri}}, \bibinfo {author} {\bibfnamefont
  {J.}~\bibnamefont {Kono}}, \bibinfo {author} {\bibfnamefont {Y.}~\bibnamefont
  {He}}, \ and\ \bibinfo {author} {\bibfnamefont {M.}~\bibnamefont {Yi}},\
  }\href {\doibase 10.1063/5.0157031} {\bibfield  {journal} {\bibinfo
  {journal} {Review of Scientific Instruments}\ }\textbf {\bibinfo {volume}
  {94}},\ \bibinfo {pages} {093902} (\bibinfo {year} {2023})}\BibitemShut
  {NoStop}%
\bibitem [{\citenamefont {Yuzbashyan}\ \emph {et~al.}(2005)\citenamefont
  {Yuzbashyan}, \citenamefont {Kuznetsov},\ and\ \citenamefont
  {Altshuler}}]{Yuzbashyan2005}%
  \BibitemOpen
  \bibfield  {author} {\bibinfo {author} {\bibfnamefont {E.~A.}\ \bibnamefont
  {Yuzbashyan}}, \bibinfo {author} {\bibfnamefont {V.~B.}\ \bibnamefont
  {Kuznetsov}}, \ and\ \bibinfo {author} {\bibfnamefont {B.~L.}\ \bibnamefont
  {Altshuler}},\ }\href {\doibase 10.1103/PhysRevB.72.144524} {\bibfield
  {journal} {\bibinfo  {journal} {Phys. Rev. B}\ }\textbf {\bibinfo {volume}
  {72}},\ \bibinfo {pages} {144524} (\bibinfo {year} {2005})}\BibitemShut
  {NoStop}%
\bibitem [{\citenamefont {Yuzbashyan}\ \emph {et~al.}(2006)\citenamefont
  {Yuzbashyan}, \citenamefont {Tsyplyatyev},\ and\ \citenamefont
  {Altshuler}}]{Yuzbashyan2006}%
  \BibitemOpen
  \bibfield  {author} {\bibinfo {author} {\bibfnamefont {E.~A.}\ \bibnamefont
  {Yuzbashyan}}, \bibinfo {author} {\bibfnamefont {O.}~\bibnamefont
  {Tsyplyatyev}}, \ and\ \bibinfo {author} {\bibfnamefont {B.~L.}\ \bibnamefont
  {Altshuler}},\ }\href {\doibase 10.1103/PhysRevLett.96.097005} {\bibfield
  {journal} {\bibinfo  {journal} {Phys. Rev. Lett.}\ }\textbf {\bibinfo
  {volume} {96}},\ \bibinfo {pages} {097005} (\bibinfo {year}
  {2006})}\BibitemShut {NoStop}%
\bibitem [{\citenamefont {Nava}\ \emph {et~al.}(2021)\citenamefont {Nava},
  \citenamefont {Rossi},\ and\ \citenamefont {Giuliano}}]{Nava2021}%
  \BibitemOpen
  \bibfield  {author} {\bibinfo {author} {\bibfnamefont {A.}~\bibnamefont
  {Nava}}, \bibinfo {author} {\bibfnamefont {M.}~\bibnamefont {Rossi}}, \ and\
  \bibinfo {author} {\bibfnamefont {D.}~\bibnamefont {Giuliano}},\ }\href
  {\doibase 10.1103/PhysRevB.103.115139} {\bibfield  {journal} {\bibinfo
  {journal} {Phys. Rev. B}\ }\textbf {\bibinfo {volume} {103}},\ \bibinfo
  {pages} {115139} (\bibinfo {year} {2021})}\BibitemShut {NoStop}%
\bibitem [{\citenamefont {Nava}\ \emph
  {et~al.}(2023{\natexlab{b}})\citenamefont {Nava}, \citenamefont {Campagnano},
  \citenamefont {Sodano},\ and\ \citenamefont {Giuliano}}]{Nava2023}%
  \BibitemOpen
  \bibfield  {author} {\bibinfo {author} {\bibfnamefont {A.}~\bibnamefont
  {Nava}}, \bibinfo {author} {\bibfnamefont {G.}~\bibnamefont {Campagnano}},
  \bibinfo {author} {\bibfnamefont {P.}~\bibnamefont {Sodano}}, \ and\ \bibinfo
  {author} {\bibfnamefont {D.}~\bibnamefont {Giuliano}},\ }\href {\doibase
  10.1103/PhysRevB.107.035113} {\bibfield  {journal} {\bibinfo  {journal}
  {Phys. Rev. B}\ }\textbf {\bibinfo {volume} {107}},\ \bibinfo {pages}
  {035113} (\bibinfo {year} {2023}{\natexlab{b}})}\BibitemShut {NoStop}%
\bibitem [{\citenamefont {Efetov}\ \emph {et~al.}(2008)\citenamefont {Efetov},
  \citenamefont {Garifullin}, \citenamefont {Volkov},\ and\ \citenamefont
  {Westerholt}}]{Efetov2008}%
  \BibitemOpen
  \bibfield  {author} {\bibinfo {author} {\bibfnamefont {K.~B.}\ \bibnamefont
  {Efetov}}, \bibinfo {author} {\bibfnamefont {I.~A.}\ \bibnamefont
  {Garifullin}}, \bibinfo {author} {\bibfnamefont {A.~F.}\ \bibnamefont
  {Volkov}}, \ and\ \bibinfo {author} {\bibfnamefont {K.}~\bibnamefont
  {Westerholt}},\ }\enquote {\bibinfo {title} {Proximity effects in
  ferromagnet/superconductor heterostructures},}\ in\ \href {\doibase
  10.1007/978-3-540-73462-8_5} {\emph {\bibinfo {booktitle} {Magnetic
  Heterostructures: Advances and Perspectives in Spinstructures and
  Spintransport}}},\ \bibinfo {editor} {edited by\ \bibinfo {editor}
  {\bibfnamefont {H.}~\bibnamefont {Zabel}}\ and\ \bibinfo {editor}
  {\bibfnamefont {S.~D.}\ \bibnamefont {Bader}}}\ (\bibinfo  {publisher}
  {Springer Berlin Heidelberg},\ \bibinfo {address} {Berlin, Heidelberg},\
  \bibinfo {year} {2008})\ pp.\ \bibinfo {pages} {251--290}\BibitemShut
  {NoStop}%
\bibitem [{\citenamefont {Mera}\ \emph {et~al.}(2018)\citenamefont {Mera},
  \citenamefont {Vlachou}, \citenamefont {Paunkovi\ifmmode~\acute{c}\else
  \'{c}\fi{}}, \citenamefont {Vieira},\ and\ \citenamefont
  {Viyuela}}]{Mera2018}%
  \BibitemOpen
  \bibfield  {author} {\bibinfo {author} {\bibfnamefont {B.}~\bibnamefont
  {Mera}}, \bibinfo {author} {\bibfnamefont {C.}~\bibnamefont {Vlachou}},
  \bibinfo {author} {\bibfnamefont {N.}~\bibnamefont
  {Paunkovi\ifmmode~\acute{c}\else \'{c}\fi{}}}, \bibinfo {author}
  {\bibfnamefont {V.~R.}\ \bibnamefont {Vieira}}, \ and\ \bibinfo {author}
  {\bibfnamefont {O.}~\bibnamefont {Viyuela}},\ }\href {\doibase
  10.1103/PhysRevB.97.094110} {\bibfield  {journal} {\bibinfo  {journal} {Phys.
  Rev. B}\ }\textbf {\bibinfo {volume} {97}},\ \bibinfo {pages} {094110}
  (\bibinfo {year} {2018})}\BibitemShut {NoStop}%
\bibitem [{\citenamefont {Zanardi}\ and\ \citenamefont
  {Paunkovi\ifmmode~\acute{c}\else \'{c}\fi{}}(2006)}]{Zanardi2006_2}%
  \BibitemOpen
  \bibfield  {author} {\bibinfo {author} {\bibfnamefont {P.}~\bibnamefont
  {Zanardi}}\ and\ \bibinfo {author} {\bibfnamefont {N.}~\bibnamefont
  {Paunkovi\ifmmode~\acute{c}\else \'{c}\fi{}}},\ }\href {\doibase
  10.1103/PhysRevE.74.031123} {\bibfield  {journal} {\bibinfo  {journal} {Phys.
  Rev. E}\ }\textbf {\bibinfo {volume} {74}},\ \bibinfo {pages} {031123}
  (\bibinfo {year} {2006})}\BibitemShut {NoStop}%
\bibitem [{\citenamefont {Quan}\ \emph {et~al.}(2006)\citenamefont {Quan},
  \citenamefont {Song}, \citenamefont {Liu}, \citenamefont {Zanardi},\ and\
  \citenamefont {Sun}}]{Zanardi2006}%
  \BibitemOpen
  \bibfield  {author} {\bibinfo {author} {\bibfnamefont {H.~T.}\ \bibnamefont
  {Quan}}, \bibinfo {author} {\bibfnamefont {Z.}~\bibnamefont {Song}}, \bibinfo
  {author} {\bibfnamefont {X.~F.}\ \bibnamefont {Liu}}, \bibinfo {author}
  {\bibfnamefont {P.}~\bibnamefont {Zanardi}}, \ and\ \bibinfo {author}
  {\bibfnamefont {C.~P.}\ \bibnamefont {Sun}},\ }\href {\doibase
  10.1103/PhysRevLett.96.140604} {\bibfield  {journal} {\bibinfo  {journal}
  {Phys. Rev. Lett.}\ }\textbf {\bibinfo {volume} {96}},\ \bibinfo {pages}
  {140604} (\bibinfo {year} {2006})}\BibitemShut {NoStop}%
\bibitem [{\citenamefont {Lu}\ and\ \citenamefont {Raz}(2017)}]{Lu2017}%
  \BibitemOpen
  \bibfield  {author} {\bibinfo {author} {\bibfnamefont {Z.}~\bibnamefont
  {Lu}}\ and\ \bibinfo {author} {\bibfnamefont {O.}~\bibnamefont {Raz}},\
  }\href {\doibase 10.1073/pnas.1701264114} {\bibfield  {journal} {\bibinfo
  {journal} {Proc. Nat. A. of Sciences}\ }\textbf {\bibinfo {volume} {114}},\
  \bibinfo {pages} {5083} (\bibinfo {year} {2017})}\BibitemShut {NoStop}%
\bibitem [{\citenamefont {B\'acsi}\ and\ \citenamefont
  {D\'ora}(2023)}]{Bacsi2023}%
  \BibitemOpen
  \bibfield  {author} {\bibinfo {author} {\bibfnamefont {A.}~\bibnamefont
  {B\'acsi}}\ and\ \bibinfo {author} {\bibfnamefont {B.}~\bibnamefont
  {D\'ora}},\ }\href {\doibase 10.1103/PhysRevB.107.125149} {\bibfield
  {journal} {\bibinfo  {journal} {Phys. Rev. B}\ }\textbf {\bibinfo {volume}
  {107}},\ \bibinfo {pages} {125149} (\bibinfo {year} {2023})}\BibitemShut
  {NoStop}%
\bibitem [{\citenamefont {Guerci}\ and\ \citenamefont
  {Nava}(2021)}]{Guerci2021}%
  \BibitemOpen
  \bibfield  {author} {\bibinfo {author} {\bibfnamefont {D.}~\bibnamefont
  {Guerci}}\ and\ \bibinfo {author} {\bibfnamefont {A.}~\bibnamefont {Nava}},\
  }\href {\doibase https://doi.org/10.1016/j.physe.2021.114895} {\bibfield
  {journal} {\bibinfo  {journal} {Physica E}\ }\textbf {\bibinfo {volume}
  {134}},\ \bibinfo {pages} {114895} (\bibinfo {year} {2021})}\BibitemShut
  {NoStop}%
\bibitem [{\citenamefont {Giuliano}\ \emph
  {et~al.}(2020{\natexlab{a}})\citenamefont {Giuliano}, \citenamefont {Nava},\
  and\ \citenamefont {Sodano}}]{Giuliano2020}%
  \BibitemOpen
  \bibfield  {author} {\bibinfo {author} {\bibfnamefont {D.}~\bibnamefont
  {Giuliano}}, \bibinfo {author} {\bibfnamefont {A.}~\bibnamefont {Nava}}, \
  and\ \bibinfo {author} {\bibfnamefont {P.}~\bibnamefont {Sodano}},\ }\href
  {\doibase https://doi.org/10.1016/j.nuclphysb.2020.115192} {\bibfield
  {journal} {\bibinfo  {journal} {Nuclear Physics B}\ }\textbf {\bibinfo
  {volume} {960}},\ \bibinfo {pages} {115192} (\bibinfo {year}
  {2020}{\natexlab{a}})}\BibitemShut {NoStop}%
\bibitem [{\citenamefont {Giuliano}\ \emph
  {et~al.}(2020{\natexlab{b}})\citenamefont {Giuliano}, \citenamefont
  {Lepori},\ and\ \citenamefont {Nava}}]{Giuliano2020a}%
  \BibitemOpen
  \bibfield  {author} {\bibinfo {author} {\bibfnamefont {D.}~\bibnamefont
  {Giuliano}}, \bibinfo {author} {\bibfnamefont {L.}~\bibnamefont {Lepori}}, \
  and\ \bibinfo {author} {\bibfnamefont {A.}~\bibnamefont {Nava}},\ }\href
  {\doibase 10.1103/PhysRevB.101.195140} {\bibfield  {journal} {\bibinfo
  {journal} {Phys. Rev. B}\ }\textbf {\bibinfo {volume} {101}},\ \bibinfo
  {pages} {195140} (\bibinfo {year} {2020}{\natexlab{b}})}\BibitemShut
  {NoStop}%
\end{thebibliography}%

\end{document}